\documentclass[default,iicol]{sn-jnl}

\usepackage{graphicx}%
\usepackage{multirow}%
\usepackage{amsmath,amssymb,amsfonts}%
\usepackage{amsthm}%
\usepackage{mathrsfs}%
\usepackage[title]{appendix}%
\usepackage{xcolor}%
\usepackage{textcomp}%
\usepackage{manyfoot}%
\usepackage{booktabs}%
\usepackage{algorithm}%
\usepackage{algorithmicx}%
\usepackage{algpseudocode}%
\usepackage{listings}%
\usepackage{cuted}
\usepackage{geometry}%
\newcommand{\asr}{Adv. Space Res.}
\newcommand{\aap}{Astron. Astrophys.}   
\newcommand{\aapr}{Astron. Astrophys. Rev.}   
\newcommand{\apj}{Astrophys. J.}   
\newcommand{\apjl}{Astrophys. J. Lett.}   
\newcommand{\jcap}{J. Cosmol. Astropart. Phys.}   
\newcommand{\jpp}{J. Plasma Phys.}
\newcommand{\mnras}{Mon. Not. R. Astron. Soc.}   
\newcommand{\nat}{Nature} 
\newcommand{\nastro}{Nat. Astron.} 
\newcommand{\pof}{Phys. Fluids}
\newcommand{\pop}{Phys. Plasmas}
\newcommand{\physrep}{Phys. Rep.}   
\newcommand{\prd}{Phys. Rev. D}   
\newcommand{\prl}{Phys. Rev. Lett.}   
\newcommand{\prx}{Phys. Rev. X}   

\makeatletter
\let\old@maketitle\@maketitle
\renewcommand{\@maketitle}{%
    \newpage\null%
    \if@remarkboxon\vbox to 0pt{\vspace*{-100pt}\hspace*{-18pt}\FMremark}\else\vskip1pt\fi%
    \hsize\textwidth\parindent0pt%
    {\hbox to \textwidth{{\Artcatfont{Accepted manuscript for Nature Astronomy, post-peer review, prior to final editorial changes. Published: \url{https://doi.org/10.1038/s41550-024-02442-1}}\hfill}}\par}%
    {\hbox to \textwidth{{\Artcatfont\ArtType\hfill}\par}}%
    \old@maketitle
}
\makeatother

\begin{document}

\title[Efficient micromirror confinement of sub-TeV cosmic rays in galaxy clusters]{Efficient micromirror confinement of sub-TeV cosmic rays in galaxy clusters}


\author*[1,2]{\fnm{Patrick} \sur{Reichherzer}}\email{patrick.reichherzer@physics.ox.ac.uk}

\author[1,3]{\fnm{Archie F.~A.} \sur{Bott}}

\author[1,4]{\fnm{Robert J.} \sur{Ewart}}

\author[1,5]{\fnm{Gianluca} \sur{Gregori}}

\author[6]{\fnm{Philipp} \sur{Kempski}}

\author[6,7]{\fnm{Matthew W.} \sur{Kunz}}

\author[1,8]{and \fnm{Alexander A.} \sur{Schekochihin}}
\affil[1]{\orgdiv{Department of Physics}, \orgname{University of Oxford}, \orgaddress{\city{Oxford} \postcode{OX1 3PU}, \country{UK}}}
\affil[2]{\orgname{Exeter College}, \orgaddress{\city{Oxford} \postcode{OX1 3DP}, \country{UK}}}
\affil[3]{\orgname{Trinity College}, \orgaddress{\city{Oxford} \postcode{OX1 3BH}, \country{UK}}}
\affil[4]{\orgname{Balliol College}, \orgaddress{\city{Oxford} \postcode{OX1 3BJ}, \country{UK}}}
\affil[5]{\orgname{Lady Margaret Hall}, \orgaddress{\city{Oxford} \postcode{OX2 6QA}, \country{UK}}}
\affil[6]{\orgdiv{Department of Astrophysical Sciences}, \orgname{Princeton University}, \orgaddress{\city{Peyton Hall, Princeton} \postcode{NJ 08544}, \country{USA}}}
\affil[7]{\orgdiv{Princeton Plasma Physics Laboratory}, \orgname{Princeton University}, \orgaddress{\city{PO Box 451, Princeton} \postcode{NJ 08543}, \country{USA}}}
\affil[8]{\orgname{Merton College}, \orgaddress{\city{Oxford} \postcode{OX1 4JD}, \country{UK}}}

\abstract{
Cosmic rays (CRs) play a pivotal role in shaping the thermal and dynamical properties of astrophysical environments, such as galaxies and galaxy clusters. Recent observations suggest a stronger confinement of CRs in certain astrophysical systems than predicted by current CR-transport theories. Here, we show that the incorporation of microscale physics into CR-transport models can account for this enhanced CR confinement. We develop a theoretical description of the effect of magnetic microscale fluctuations originating from the mirror instability on macroscopic CR diffusion. We confirm our theory with large-dynamical-range simulations of CR transport in the intracluster medium (ICM) of galaxy clusters and kinetic simulations of CR transport in micromirror fields. We conclude that sub-TeV CR confinement in the ICM is far more effective than previously anticipated on the basis of Galactic-transport extrapolations. The transformative impact of micromirrors on CR diffusion provides insights into how microphysics can reciprocally affect macroscopic dynamics and observable structures across a range of astrophysical scales.
}


\maketitle
A good theory of cosmic-ray (CR) transport is crucial for advancing our understanding of phenomena in the Universe, including the formation and evolution of galaxies and galaxy clusters. CRs, through their transport characteristics, not only influence their environment but also modulate their own (re)acceleration and confinement efficiency as well as the observable photon and neutrino emission.

The transport characteristics of CRs in magnetic-field structures depend on the scattering efficiency and mechanism, both of which are influenced by the properties of the ambient plasma. Specifically, within a weakly collisional, high-$\beta$ plasma, i.e., one in which the thermal pressure much exceeds the magnetic pressure, deviations from local thermodynamic equilibrium provide free energy for fast-growing Larmor-scale instabilities, leading to distortions in magnetic fields on thermal-ion kinetic scales. In such a high-$\beta$ plasma, two characteristic scales are relevant for describing the global transport of CRs: the macroscale of the magnetic turbulence, characterized by the correlation length $l_\mathrm{c}$ or the ``Alfvén scale'' $l_\mathrm{A}$, and the microscale $l_\mathrm{mm}$ of the micromirrors created by the mirror instability \citep{Schekochihin_2005ApJ...629..139S, Kunz_2022hxga.book...56K}. The prefix “micro” refers to scales much smaller than the ${\gtrsim} \mu\mathrm{pc}$ gyroradii of ${\gtrsim}  100\,\mathrm{MeV}$ CRs, and serves to distinguish the plasma-kinetic-scale “micromirrors” ($_\mathrm{mm}$) from the large-scale magnetic mirrors that also influence CR transport \citep{Cesarsky_1973ApJ...185..153C, Reichherzer_2020MNRAS.498.5051R, Lzarian_2021ApJ...923...53L}. While there are other micro-instabilities, the magnetic fluctuations created by the mirror instability are stronger and thus more influential for CR transport \citep{Kunz_2014PhRvL.112t5003K,Bott_2023}.  
The physics associated with these micro- and macroscales introduces three distinct transport regimes, which depend on the CR energy.

First, in the high-energy ($_\mathrm{he}$) limit, CRs with gyroradii $r_\mathrm{g} \gg l_\mathrm{c}$ (energies $E\gtrsim 100\,$EeV for typical turbulence-driving scales in galaxy clusters) undergo scattering by small angles of the order $\delta \Theta \sim l_\mathrm{c}/r_\mathrm{g}$ at characteristic times~$\delta t \sim l_\mathrm{c}/c$, leading to a scattering rate $\nu_\mathrm{he} \sim \delta \Theta^2/\delta t \sim c l_\mathrm{c}/r_\mathrm{g}^2$ and, therefore, to a CR diffusion coefficient~$\kappa_\mathrm{he} \sim c^2/\nu_\mathrm{he} \sim cr_\mathrm{g}^2/l_\mathrm{c} \propto E^2/l_\mathrm{c}$ \citep[e.g.,][]{Aloisio_2004ApJ...612..900A}. This scaling is indeed observed both in numerical simulations \citep[e.g.,][]{Kotera_2008PhRvD..77b3005K, Subedi_2017ApJ, Reichherzer_2022SNAS....4...15R} and (scaled) laboratory experiments \citep{Chen_2020ApJ} and serves as input for propagation models of ultra-high-energy CRs in galaxy clusters \citep{Condorelli_2023impact}.

As the high-energy regime is, thus, believed to be understood, recent studies of CR transport have predominantly focused on the second, mesoscale regime, $ r_\mathrm{g} \lesssim l_\mathrm{c}$, in which CRs scatter resonantly off inertial-range turbulent structures \citep{Jokipii1966, Giacalone_1999ApJ...520..204G, Lemoine_2023arXiv230403023L, Kempski_2023arXiv230412335K}. This second regime is likely relevant for explaining CR spectra \citep{Amato_2018AdSpR..62.2731A}, e.g., the steepening of the CR spectrum in the Galactic Center from GeV to PeV energies \citep{Reichherzer_2022SNAS....4...15R, Luque_2023A&A...672A..58D, Cao_2023arXiv230505372C, Zhang_2023arXiv230506948Z}. 

We argue, and confirm numerically, that in high-$\beta$ plasmas, the presence of microstructures caused by plasma instabilities introduces a third regime, whose physics is similar to that of the first, but with the micromirror scale $l_\mathrm{mm}$ playing the role of $l_\mathrm{c}$ and the requirement that~$r_\mathrm{g} \gg  l_\mathrm{mm}$ ($E\gg 100\,$MeV). We show that this microscale physics largely overrides the mesoscale resonant scattering and streaming. We apply our theory to the intracluster medium (ICM), a representative high-$\beta$ plasma, and determine the transition between micro- and macrophysics-dominated transport to be at TeV energies, only weakly influenced by mesoscale physics. We confirm this theory with a novel method (described in Section~\ref{sec:method}) that incorporates the microscales~($l_\mathrm{mm} \sim 100\,$npc), the macroscales ~($l_{\rm c} \sim 100\,$kpc), and the vast range in between.

\section{Results}\label{sec2}
\subsection{Effect of micromirrors on large-scale CR transport}\label{sec:micromirror}
It has long been realized that plasma instabilities may dominate the transport of low-energy CRs. However, the instabilities most often highlighted in the literature arise from the CRs themselves, rather than from the thermal plasma. One prominent example, especially for Galactic transport of CRs below $100\,$GeV, is the streaming instability \citep{Kulsrud_1969ApJ...156..445K}. This instability generates fluctuations in the magnetic field that in turn scatter the CRs and thereby reduce their streaming velocity to be comparable to the Alfv\'{e}n speed in the plasma \citep{Kulsrud_1969ApJ...156..445K, Skilling_1971ApJ...170..265S, Blasi_2012PhRvL.109f1101B, Shalaby_2023arXiv230518050S}. In high-$\beta$, weakly collisional plasmas, there also exists a variety of instabilities that are driven by pressure anisotropies and generate magnetic fluctuations on ion-Larmor scales \citep[see][and references therein]{Kunz_2022hxga.book...56K}. 
The pressure anisotropies arise from the (approximate) conservation of particles' adiabatic invariants during the local stretching and compression of magnetic fields \citep{CGL_1956}. 
In the present context, the mirror instability \citep{Barnes_1966PhFl, Hasegawa_1969PhFl} is of particular interest because its saturated amplitude $\delta B_\mathrm{mm} \sim B/3$ is of the same order of magnitude as the ambient magnetic field $B$ \citep{Kunz_2014PhRvL.112t5003K, Riquelme_2015ApJ...800...27R, Ley_2023arXiv230916751L}. The other well-known instability arising in such plasmas, the firehose instability, is unlikely to affect CR transport because its expected saturation amplitude is small under ICM conditions: $\delta B_{\rm f} \sim (\tau \Omega_{i})^{-1/4} B$~\citep{Kunz_2014PhRvL.112t5003K,Bott_2023}, where $\tau$ is the timescale over which a firehose-susceptible plasma evolves macroscopically, and $\Omega_i \sim 0.01\, (B/3\,\mu\mathrm{G})\,\mathrm{s}^{-1}$ is the non-relativistic thermal-ion gyrofrequency. In the ICM, $\tau \Omega_i \sim 10^{11}$ \citep{Rosin_2011MNRAS.413....7R}, so $\delta B_{\rm f} \sim 10^{-3} B$. By analogy to~(\ref{eq:nu_mm}), it follows that the CR scattering rate $\nu_{\rm f}$ of CRs off firehoses in the ICM is much smaller than that off the mirrors: $\nu_{\rm f}/\nu_{\rm mm} \sim 10^{-7}$, with the firehose scale taken to be comparable to the thermal-ion gyroradius~\citep{Kunz_2014PhRvL.112t5003K,Bott_2023}.

To determine the impact of these micromirrors on CR transport, we begin by working out the relevant theoretical predictions for diffusion coefficients of CRs scattering at such strong fluctuations. Note that a previous assumption of weaker micromirror fluctuations led to a different, much larger diffusion coefficient based on calculations using quasi-linear theory \citep{Hall_1981MNRAS.197..977H}.
The velocity change~$\delta \boldsymbol{v}$ of a relativistic CR with gamma factor $\gamma = (1-v^2/c^2)^{-1/2}$, charge $q=Ze$, and mass $m$ in a magnetic-field structure of scale~$l_\mathrm{mm} \ll r_\mathrm{g}$ and vector amplitude~$\delta \boldsymbol{B}_\mathrm{mm}$ is given in the small-angle limit~$|\delta \boldsymbol{v}| \ll v$ by integrating the equation of motion $\gamma m \,\mathrm{d}\boldsymbol{v}/\mathrm{d}t = q\, (\boldsymbol{v}/c)\boldsymbol{\times} \delta \boldsymbol{B}_\mathrm{mm}$ along the CR path:
\begin{equation}\label{eq:delta_v_vec}
    \delta \boldsymbol{v} \sim \frac{q}{\gamma m c v} \int_{0}^{l_\mathrm{mm}} \mathrm{d}l\,\boldsymbol{v} \boldsymbol{\times} \delta \boldsymbol{B}_\mathrm{mm}
    \,.
\end{equation}
Assuming relativistic CRs with $v \approx c$, energy $E=\gamma m c^2$, and gyroradius $r_\mathrm{g} = \gamma m c^2/qB$ determined by the ambient magnetic field $B \gtrsim \delta B_\mathrm{mm}$, the scattering angle at characteristic times~$\delta t~\sim~l_\mathrm{mm}/c$ is
\begin{equation}\label{eq:delta_v}
    \delta \Theta \sim \frac{|\delta \boldsymbol{v}|}{c} \sim  \frac{l_\mathrm{mm}}{r_\mathrm{g}} \frac{\delta B_\mathrm{mm}}{B} 
    \,.
\end{equation}
Assuming that these small-angle deflections add up as a correlated random walk, the scattering rate is
\begin{equation}\label{eq:nu_mm}
    \nu_\mathrm{mm} \sim \frac{\delta \Theta^2}{\delta t} \sim \frac{c\,l_\mathrm{mm}}{r_\mathrm{g}^2} \left(\frac{\delta B_\mathrm{mm}}{B} \right)^{2}\,, 
\end{equation}
which implies a spatial diffusion coefficient of
\begin{equation}\label{eq:kappa_mm}
    \kappa_\mathrm{mm} \sim \frac{c^2}{\nu_\mathrm{mm}} \sim \frac{c\,r_\mathrm{g}^2}{l_\mathrm{mm}} \left(\frac{\delta B_\mathrm{mm}}{B} \right)^{-2} \propto E^2 l_\mathrm{mm}^{-1}\,. 
\end{equation}
As usual, more energetic CRs diffuse much faster.

\begin{figure}
\centering
\includegraphics[width=0.98\columnwidth] {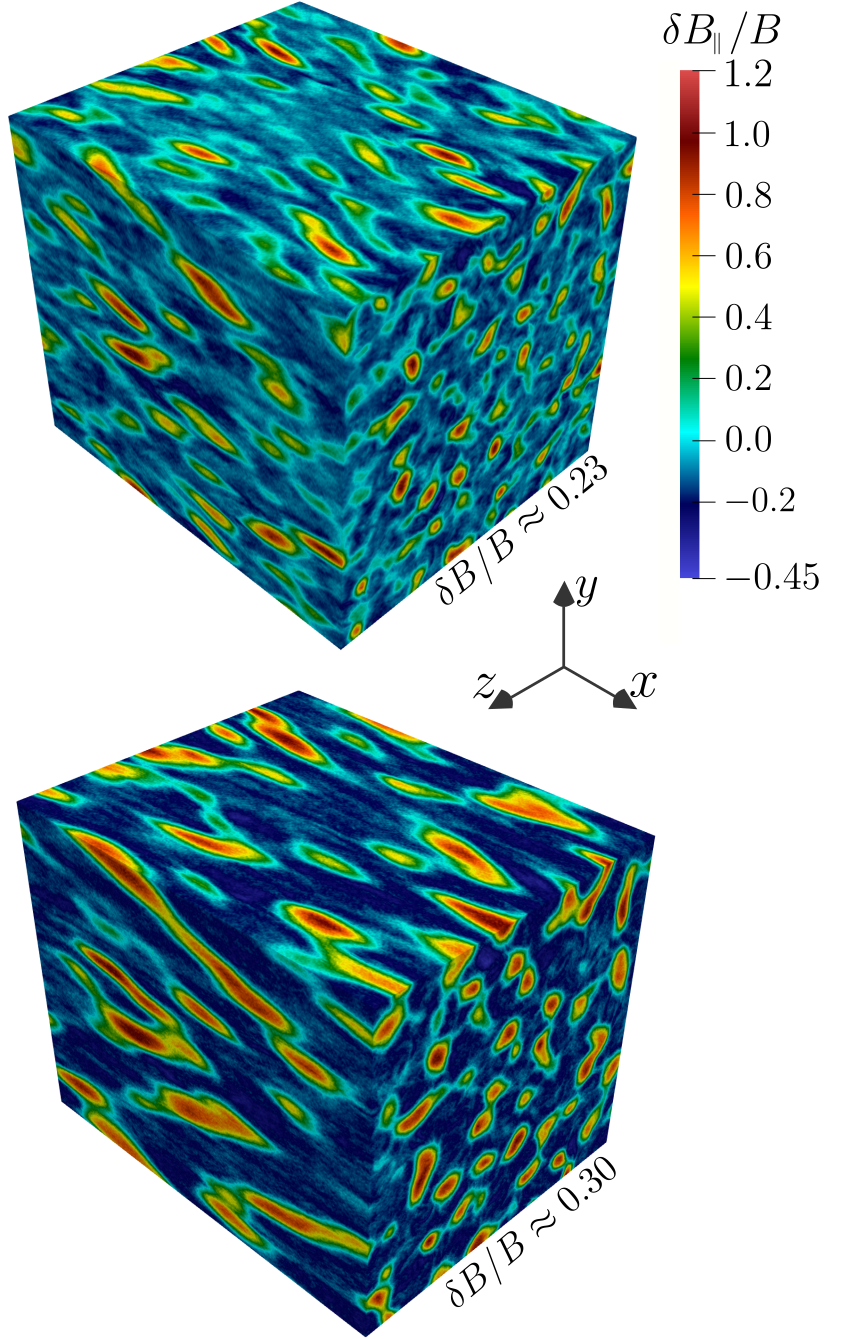}
\vspace*{-0.2cm} 
\caption{\textbf{Micromirror field generated by the PIC simulation.} Color shows fluctuations $\delta B_\parallel$ along the field $B$, which is aligned with the $x$-axis. We show two snapshots of the 3D field during its secular evolution, characterized by different $\delta B_\mathrm{rms}/B$, as indicated in the plots (the right snapshot is later in the evolution).
\label{fig:mirrorfield}}
\end{figure}

In arriving at~(\ref{eq:kappa_mm}), we effectively assumed that micromirrors are described by only one characteristic scale,~$l_\mathrm{mm}$. In reality, micromirrors are anisotropic (see ellipsoid-like shapes in Figure~\ref{fig:mirrorfield}) with scales perpendicular ($\perp$) and parallel ($\parallel$) to the ambient magnetic field that satisfy $l_{\perp, \mathrm{mm}} \ll l_{\parallel, \mathrm{mm}}$. 
While gyrating through this field, CRs with $r_\mathrm{g} \gg l_\mathrm{mm}$ mostly traverse micromirrors perpendicularly. Only low-energy CRs satisfying~$r_\mathrm{g}\, v_\perp/c \lesssim l_{\perp, \mathrm{mm}} \ll l_{\parallel, \mathrm{mm}}$ are an exception and should be treated analogously to thermal electrons with negligible $r_\mathrm{g}$ getting scattered \citep{Riquelme_2015ApJ...800...27R, Zhou_2023} and trapped \citep{Komarov_2016MNRAS.460..467K} in the micromirrors. This subpopulation makes a negligible contribution to the overall transport of CRs with $r_\mathrm{g} \gg l_{\perp, \mathrm{mm}}$ considered here.

For $r_\mathrm{g} \gg l_{\perp, \mathrm{mm}}$, CRs will sample many different micromirrors, with deflections adding up as a correlated random walk. During one gyro-orbit, CRs will travel $\Delta l_\parallel \sim 2\pi r_\mathrm{g}\,v_\parallel/c$ in the field-parallel direction. CRs with large pitch angles satisfying $\Delta l_\parallel \lesssim l_\mathrm{\parallel, mm}$, i.e., $v_\perp / v_\parallel \sim c / v_\parallel \gtrsim 2\pi r_\mathrm{g}/l_\mathrm{\parallel, mm}$, that sample the same micromirror repeatedly, may become relevant only at low energies $r_\mathrm{g}\lesssim l_\mathrm{\parallel, mm}/2\pi$, not considered in this study. The scattering rate associated with the parallel micromirror perturbation $\delta B_\parallel \sim B_\mathrm{mm}$ decreases with decreasing pitch angle, but this is overcome by scattering at the perpendicular micromirror component~$\delta B_\perp\sim \delta B_\parallel l_{\perp, \mathrm{mm}} / l_{\parallel, \mathrm{mm}}$ for~$v_\perp/v_\parallel \lesssim l_{\perp, \mathrm{mm}}/l_{\parallel, \mathrm{mm}} \ll 1$. Except for this cone containing CRs with small pitch angles, from which they escape quickly on the time scale $t_\mathrm{esc} \sim \nu_\mathrm{mm}^{-1} \, l_{\perp, \mathrm{mm}}/ l_{\parallel, \mathrm{mm}}$, gyrating CRs cross micromirrors perpendicularly faster than they traverse them in the parallel direction, implying~$l_\mathrm{mm} \sim l_{\perp, \mathrm{mm}}$ to be the relevant scale. 

For application to the ICM, we estimate
$l_{\mathrm{mm}}$ using an asymptotic theory of the mirror instability's nonlinear evolution \citep{Rincon_2015MNRAS.447L..45R} supported by previous numerical studies \citep{Kunz_2014PhRvL.112t5003K, Melville_2016MNRAS.459.2701M}: $l_{\mathrm{mm}} \sim (\tau \Omega_i)^{1/8} r_{\mathrm{g},i}$, where $\tau$ is the timescale over which a micromirror-susceptible plasma evolves macroscopically. Applying the theory to an ICM \citep{Govoni_2017A&A...603A.122G, Kunz_2022hxga.book...56K} with $B\sim 3\,\mu$G, the thermal-ion gyroradius $r_{\mathrm{g},i} \sim (2T/m_i)^{1/2}/\Omega_i \sim 1\,$npc, the mean galaxy cluster temperature $T\sim 5\,$keV, the thermal-ion mass $m_i$, and $\tau\sim10^{12}\,\mathrm{s}$ \citep{Rosin_2011MNRAS.413....7R, Rincon_2015MNRAS.447L..45R} yields~$l_\mathrm{mm} \sim l_{\perp, \mathrm{mm}} \sim 100\,r_{\mathrm{g},i} \sim 100\,$npc \citep{Melville_2016MNRAS.459.2701M}, only a factor of few smaller than the gyroradius of a GeV CR. In combination with~(\ref{eq:kappa_mm}), this gives us the estimate
\begin{equation}\label{eq:kappa}
    \begin{split}
    \kappa_\mathrm{mm} &\sim 10^{30} \,Z^{-2}\,\left(\frac{l_\mathrm{mm}}{100\,\mathrm{npc}} \right)^{-1}\left(\frac{B}{3\,\mu\mathrm{G}} \right)^{-2}\\
    &~~~~~~~~~~~~~\times \left(\frac{\delta B_\mathrm{mm}/B}{1/3} \right)^{-2} \left(\frac{E}{\mathrm{TeV}} \right)^2\,\mathrm{cm}^2\,\mathrm{s}^{-1}, \\
    &\sim 10^{30} \,Z^{-2}\,\left(\frac{T}{5\,\mathrm{keV}} \right)^{-1/2}\left(\frac{B}{3\,\mu\mathrm{G}} \right)^{-1}\\
    &~~~~~~~~~~~~~\times \left(\frac{\delta B_\mathrm{mm}/B}{1/3} \right)^{-2} \left(\frac{E}{\mathrm{TeV}} \right)^2\,\mathrm{cm}^2\,\mathrm{s}^{-1}
    \,.
    \end{split}
\end{equation}
This estimate is valid provided $l_\mathrm{mm} \ll r_\mathrm{g}$ and $\delta t \,\nu_\mathrm{mm}  \sim (l_\mathrm{mm}/r_\mathrm{g})^2 (\delta B_\mathrm{mm}/B)^2 \ll 1$ ($E \gg 100\,$MeV).

We now show that the diffusion coefficient~(\ref{eq:kappa}) is associated with parallel transport along field lines by demonstrating that the perpendicular diffusion coefficient is negligible. Each scattering at characteristic times~$\delta t~\sim~l_\mathrm{mm}/c$ moves the gyrocenter by a distance $\Delta r_\perp \sim r_g \delta \Theta$ in the plane perpendicular to the local magnetic field line. Using the estimate~(\ref{eq:delta_v}) for the scattering angle leads to the perpendicular diffusion coefficient
\begin{equation}\label{eq:kappa_perp}
    \begin{split}
    \kappa_{\perp, \mathrm{mm}} &\sim \frac{\Delta r_\perp^2}{\delta t} \sim \frac{r_g^2 \delta \Theta^2}{l_\mathrm{mm}/c} \sim c l_\mathrm{mm} \left(\frac{\delta B_\mathrm{mm}}{B} \right)^{2} \\
    &\sim \frac{l_\mathrm{mm}^2}{r_\mathrm{g}^2} \left(\frac{\delta B_\mathrm{mm}}{B} \right)^{4} \kappa_\mathrm{mm}\,.
    \end{split}
\end{equation}
Since $\kappa_{\perp, \mathrm{mm}} \ll \kappa_\mathrm{mm}$ for $r_\mathrm{g} \gg l_\mathrm{mm}$ and $\delta B_\mathrm{mm} \lesssim B$, it is the parallel diffusion $\kappa_{\parallel, \mathrm{mm}} \sim \kappa_\mathrm{mm}$ along field lines that dominates. The smaller perpendicular diffusion coefficients arise from anisotropic scattering. The degree to which this anisotropy enhances the parallel diffusion coefficient depends on the specifics of pitch-angle scattering and the properties of $B_\perp$, both of which warrant further investigation.

\begin{figure}
\centering
\includegraphics[width=1.02\columnwidth] {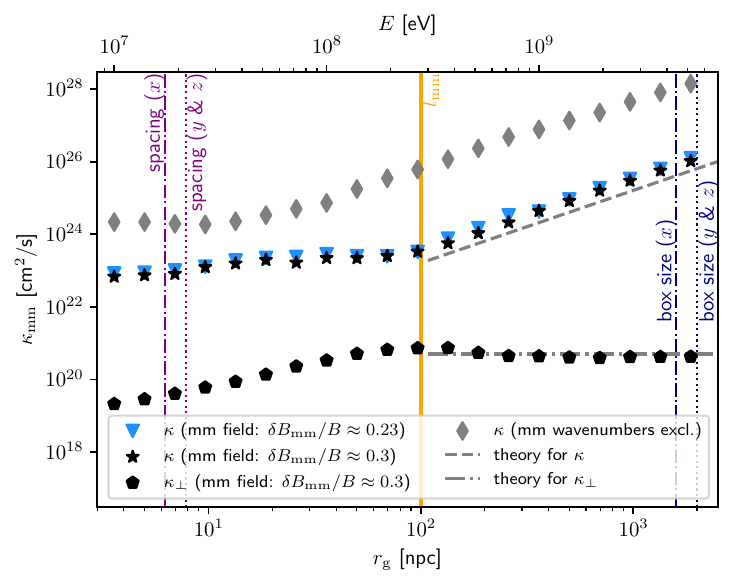}
\vspace*{-0.5cm} 
\caption{\textbf{Diffusion coefficients of CRs in micromirror fields}. Our predictions~(\ref{eq:kappa}) (gray dashed line) and (\ref{eq:kappa_perp}) (gray dash-dotted line), valid for $l_\mathrm{mm} \ll r_\mathrm{g}$, agree well with the computed diffusion coefficients. The diffusion coefficient $\kappa$ is dominated by the parallel diffusion, the perpendicular diffusion coefficient $\kappa_\perp$ is negligible. Note that $\kappa$ decreases with $\delta B_\mathrm{mm}/B$ (triangles vs. stars). We include results for CRs with gyroradii smaller than the micromirror scale $l_\mathrm{mm}$ (vertical orange line) to highlight the change in transport regimes, which follows from our theory. The gray markers show the transport of CRs through the residual field that results after filtering out the wavenumbers associated with the micromirrors, leaving only numerical noise. The purple vertical (dash-)dotted lines show grid resolution and box sizes along the three axes. $Z=1$ is used for the energy scale.
\label{fig:mirror_kappa}}
\end{figure}

To validate our theoretical prediction for the diffusion coefficients~(\ref{eq:kappa}) and~(\ref{eq:kappa_perp}), we performed a numerical experiment in which a spectrum of CRs was integrated in a magnetic field containing micromirrors generated self-consistently via a particle-in-cell (PIC) simulation (Section~\ref{sec:mm_sim}). For our numerical experiment, we selected two representative realizations of the 3D field during its secular evolution, visualized in Figure~\ref{fig:mirrorfield}. We then determined the diffusion coefficients of the CRs in both fields by integrating the CR equation of motion. The results are shown in Figure~\ref{fig:mirror_kappa}. The diffusion coefficients in the micromirror fields show good agreement with~(\ref{eq:kappa}) and~(\ref{eq:kappa_perp}).

\subsection{The micro-macrophysics transition is at TeV CR energies}\label{sub:transition}
In Section~\ref{sec:micromirror}, we derived the diffusion coefficient $\kappa_\mathrm{mm} \propto l_\mathrm{mm}^{-1} E^2$ associated with CR scattering at micromirrors of scale $l_\mathrm{mm}$. 
To determine the upper bound for the energies at which the scattering off micromirrors dominates CR transport, one needs a model of the competing contribution to it from the resonant scattering off mesoscale magnetic turbulence.
In Section~\ref{sec:mesoscales}, we present models of CR diffusion based on resonant scattering in the (mesoscale) inertial range of magnetic turbulence stirred at the macroscale $l_\mathrm{c}$, leading to the diffusion coefficient 
\begin{equation}\label{eq:kappa_res}
    \kappa_\mathrm{res} \sim c\,l_\mathrm{c} \left(  \frac{r_\mathrm{g}}{l_\mathrm{c}} \right)^{\delta} \propto E^{\delta}l_\mathrm{c}^{-\delta+1}\,,
\end{equation}
where the model-dependent exponent is $0 \leq \delta \leq 1/2$. The mechanism with the smallest diffusion coefficient dominates CR transport. The transition between the micromirror and resonant-scattering transport regimes occurs when $\kappa_\mathrm{mm} \sim \kappa_\mathrm{res}$. Equating (\ref{eq:kappa_mm}) and (\ref{eq:kappa_res}) determines the gyroradius corresponding to this transition:
\begin{equation}\label{eq:r_g_tran}
    r_\mathrm{g} \sim l_\mathrm{c} \left(\frac{\delta B_\mathrm{mm}}{B}\right)^{2/(2-\delta)}  \left(\frac{l_\mathrm{mm}}{l_\mathrm{c}}\right)^{1/(2-\delta)}\,.
\end{equation}

This translates into a $\delta$-dependent estimate for the transition energy:

\begin{align}\label{eq:energy}
\begin{split}
   E &\sim  Z\,\left(\frac{B}{3\,\mu\mathrm{G}}\right) \left(\frac{l_\mathrm{c}}{100\,\mathrm{kpc}}\right) 
     \times\\
    &\begin{cases}
      \displaystyle 300 \,\left(\frac{\delta B_\mathrm{mm}}{B}\right)^{2/(2-\delta)}\left(\frac{l_\mathrm{mm}}{l_\mathrm{c}}\right)^{1/(2-\delta)} \, \mathrm{EeV}\,, & \text{general $\delta$}\,,\\
      \displaystyle 5\, \left(\frac{\delta B_\mathrm{mm}/B}{1/3}\right)^{6/5}\left(\frac{l_\mathrm{mm}/l_\mathrm{c}}{10^{-12}}\right)^{3/5}\, \mathrm{TeV}\,, & \delta = 1/3\,,\\
      \displaystyle 600\,\left(\frac{\delta B_\mathrm{mm}/B}{1/3}\right)^{4/3}\left(\frac{l_\mathrm{mm}/l_\mathrm{c}}{10^{-12}}\right)^{2/3}\, \mathrm{GeV}\,, & \delta = 1/2\,.
    \end{cases} 
\end{split}
\end{align}

Below this energy, magnetic micromirrors dominate CR diffusion. The factor involving the ratio $l_\mathrm{mm}/l_\mathrm{c}$ accounts for the scale separation between micro- and macrophysics, which is ${\sim}10^{-12}$ in our fiducial ICM under the same assumptions as in Section~\ref{sec:micromirror}. 

\begin{figure*}[!t]
\hspace*{-0.6cm} 
\centering
\includegraphics[width=2.15\columnwidth] {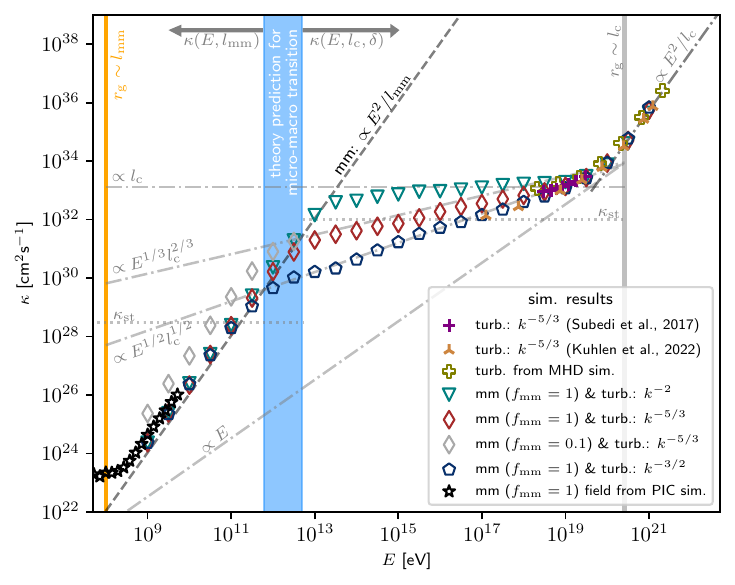}
\vspace*{-0.6cm} 
\caption{\textbf{Diffusion coefficients of cosmic rays (CRs) in the ICM as functions of CR energy.} The black stars show the diffusion coefficients in the micromirror field generated by a PIC simulation (Section~\ref{sec:mm_sim}). The olive empty crosses show the diffusion coefficients of CRs in MHD turbulence without a guide field (Section~\ref{sec:mhd_sim}). The other empty markers show the diffusion coefficients computed in isotropic synthetic turbulence with a large inertial range (Section~\ref{sec:synthetic_turbulence}), together with our stochastic micromirror-scattering model (Section~\ref{sec:mathod_mmm}), assuming the volume-filling fraction $f_\mathrm{mm}$ of micromirrors indicated in the legend (see Section~\ref{sub:intermittency}). The grey dash-dotted lines represent theories for CR transport depending on the macroscale $l_\mathrm{c}$ according to~(\ref{eq:kappa_res}), including the most efficient (Bohm) and the least efficient (energy-independent) diffusion scenarios. The grey dotted line represents the diffusion due to streaming instability, according to~(\ref{eq:kappa_st}) in Section~\ref{sec:streaming}. The gray dashed line represents our prediction of the diffusion due to micromirrors according to~(\ref{eq:kappa}). The vertical light-blue bar indicates our estimate~(\ref{eq:energy}) for the micro-macro transition for the most likely range of $\delta$ between $1/3$ and $1/2$. Simulation results from \cite{Subedi_2017ApJ} and \cite{Kuhlen_2022arXiv221105881K} illustrate the best resolution towards the limit~$r_\mathrm{g} \ll l_\mathrm{c}$ achieved prior to the present results with synthetic turbulence on a grid ($2048^3$ grid points) and nested grids, respectively. The effects of field-line tangling are not considered, which is expected to reduce the global CR diffusion coefficients by a factor of three.}
\label{fig:result}
\end{figure*}

To test this prediction, we performed numerical simulations of CR transport in the ICM (detailed in Section~\ref{sec:method}), modeling the effects of both the micromirrors (Section~\ref{sec:mathod_mmm}) and of the turbulent cascade up to~$l_{\rm c} \sim 100\,$kpc (Section~\ref{sec:mesoscales}). Figure~\ref{fig:result} summarizes our results by presenting the CR diffusion coefficient as a function of energy. The vertical light-blue bar indicates our estimate (\ref{eq:energy}) for the micro-macro transition assuming the most likely range of $\delta \in[1/3,1/2]$. While this estimate of the micro-macrophysics transition at TeV CR energies is numerically confirmed using synthetic turbulence, we also used magnetic fields from direct PIC and MHD simulations to validate the consistency of our numerical approach at micro- and macroscales, respectively, and capture all relevant diffusion coefficients discussed in the literature, detailed as points~(i) to (iii) in Section~\ref{sec:mesoscales}.

\begin{figure*}[!t]
\centering
\includegraphics[width=2.08\columnwidth] {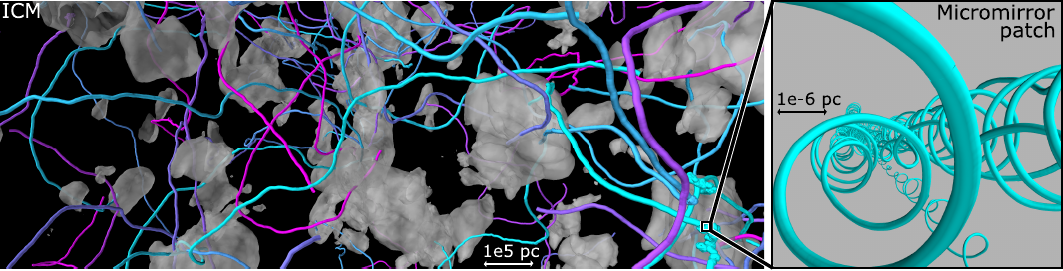}
\vspace*{-0.5cm} 
\caption{\textbf{Visualization of example CR trajectories through spatially intermittent micromirror patches.} The numerical experiment to study the effective CR transport in a two-phase medium is described in Section~\ref {sec:mathod_two_phase_medium}. The grey surfaces are the isosurfaces of a threshold field strength~$B_\mathrm{>}$. In our simplified numerical experiment, these isosurfaces are assumed to enclose the micromirror patches, inside which the diffusion coefficient is much smaller than outside. Therefore, a given choice of~$B_>$ corresponds to a certain value of $f_\mathrm{mm}$ (see Section~\ref{sec:mathod_two_phase_medium}). Example CR trajectories show increased deflections within the micromirror patches (see, e.g., lower right corner of the left panel and the zoom into a micromirror patch in the right panel). Taking the patches to be static is suitable for small $f_\mathrm{mm}$ as demonstrated in Section~\ref {sec:mathod_two_phase_medium}.
\label{fig:visualization}}
\end{figure*}

\subsection{Case of spatially intermittent micromirrors}\label{sub:intermittency}
Thus far, we have effectively assumed that the micromirrors permeate the plasma uniformly. In reality, the situation is more complicated: micromirrors will most likely appear in spatially intermittent and temporally transient patches wherever turbulence leads to local amplification of the magnetic field at a rate that is sufficiently large to engender positive pressure anisotropy exceeding the mirror-instability threshold \citep[${\sim}1/\beta$; see, e.g,][and references therein]{Melville_2016MNRAS.459.2701M,Squire_2023arXiv230300468S}. This gives rise to an effectively two-phase plasma (see Figure~\ref{fig:visualization} for an illustration), with two different effective scattering rates: $\nu_\mathrm{mm} \sim c^2/\kappa_\mathrm{mm}$ in micromirror patches and $\nu_\mathrm{res} \sim c^2/\kappa_\mathrm{res}$ elsewhere (instead of $\kappa_\mathrm{res}$, one could also use $\kappa_\mathrm{st}$ associated with the streaming instability -- see Section~\ref{sec3}). For the purpose of modeling CR scattering in such a plasma, we introduce the effective micromirror fraction $f_\mathrm{mm}$ to quantify the probability of CR being scattered by the micromirrors.

By definition, the effective scattering rate in a two-phase medium is \citep{Kalnin_1998JPhA...31.7227K, Kotera_2008PhRvD..77l3003K}
\begin{equation}\label{eq:nu_eff_f}
    \nu_\mathrm{eff} = f_\mathrm{mm}\, \nu_\mathrm{mm}  + (1-f_\mathrm{mm})\, \nu_\mathrm{res}\,.
\end{equation}
The effective diffusion coefficient is then
\begin{equation}\label{eq:kappa_eff}
    \kappa_\mathrm{eff} \sim \frac{\kappa_\mathrm{mm}}{f_\mathrm{mm}  + (1-f_\mathrm{mm}) \kappa_\mathrm{mm}/\kappa_\mathrm{res}}\,.
\end{equation}
The transition at which micromirror transport takes over from resonant scattering is entirely independent of~$f_\mathrm{mm}$: $\kappa_\mathrm{eff} \sim\kappa_\mathrm{res}$ when~$\kappa_\mathrm{mm} \sim \kappa_\mathrm{res}$. However, the asymptotic scaling~$\kappa_\mathrm{eff} \sim \kappa_\mathrm{mm} / f_\mathrm{mm}$ is only reached at CR energies for which~$\kappa_\mathrm{mm} / \kappa_\mathrm{res} \lesssim f_\mathrm{mm}/(1-f_\mathrm{mm})$, pulling the transition energy down by a factor of~$f_\mathrm{mm}^{1/(2 - \delta)}$. This is not a very strong modification of our cruder ($f_\mathrm{mm} = 1$) estimate~(\ref{eq:energy}) unless $f_\mathrm{mm}$ is extremely small.

The most intuitive interpretation of $f_\mathrm{mm}$ is that it is the fraction of the plasma volume occupied by the micromirrors. This, however, requires at least two caveats: (i) the lifetime of micromirror patches, determined by the turbulent dynamics, can be shorter than the time for a CR to diffuse through the patch; (ii) if the micromirror patches form solid macroscopically extended three-dimensional blobs, it is possible to show that CRs typically do not penetrate much farther than $\lambda_\mathrm{mm}$ into the patches. This leaves the patch volume largely uncharted (see Appendix~3 of~\cite{Kotera_2008PhRvD..77l3003K}). The second concern obviates the first (see Section~\ref{sec:mathod_two_phase_medium}). Under such a scenario, the effective CR diffusion in the ICM might be determined primarily by such factors as the typical size of the patches and the distance between them \citep{Butsky_2023arXiv230806316B}. However, the scenario of micromirror-dominated transport is made more plausible as the diffusion is mostly along the field lines. In this one-dimensional problem, CRs bounce between mirror patches on the same field line until they have a lucky streak in diffusion and pass directly through a micromirror patch. This trapping effect leads to efficient confinement, if the influence of potential field-line separation and cross-field diffusion is found to be negligible, though this remains subject to further research.

\begin{figure}
\centering
\includegraphics[width=1.03\columnwidth] {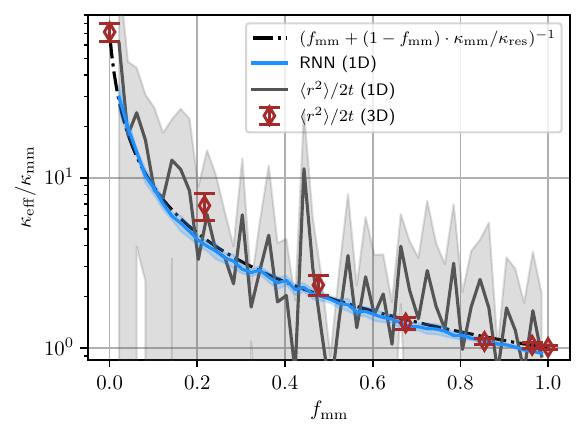}
\vspace*{-0.55cm} 
\caption{\textbf{Effective diffusion coefficient of CRs in a two-phase medium vs. the effective micromirror fraction $f_\mathrm{mm}$.} We model the CR transport through a two-phase medium in 3D and 1D (to reduce simulation costs), as explained in Section~\ref{sec:mathod_two_phase_medium}. The 3D case (red diamonds) shows the diffusion coefficient of $300\,$GeV CRs computed at their trajectory lengths of ${\simeq}10 l_\mathbf{c}$. The blue and black lines show the 1D results on 400 test trajectories using our recurrent neural network (RNN) and the classical analysis method. Error bars for the 3D data represent the standard deviation (SD) across different realizations of patches for each $f_\mathrm{mm}$, while for the 1D cases, the colored contours show the SD across 400 CRs from a single realization. The dash-dotted black line represents expected values for averaged scattering frequencies of CRs in the two-phase plasma, formally expressed in~(\ref{eq:kappa_eff}). The diffusion coefficients $\kappa_\mathrm{mm}$ and $\kappa_\mathrm{res}$ are recovered for $f_\mathrm{mm} = 1$ and $f_\mathrm{mm} = 0$, respectively.
\label{fig:result_fmm}}
\end{figure}

Our model formula~(\ref{eq:kappa_eff}) proves to be a good prediction even in a simple modification of our numerical experiment with synthetic fields, designed to model micromirror patches (see Section~\ref{sec:mathod_two_phase_medium}). Its results are shown in Figure~\ref{fig:result_fmm}. It is a matter of future work to determine the precise dependence of $f_\mathrm{mm}$ on the morphology and dynamics of magnetic fields and micromirror-unstable patches in high-$\beta$ turbulence -- itself a system that has only recently become amenable to numerical modeling \citep{Arzamasskiy_2023PhRvX..13b1014A, Squire_2023arXiv230300468S, Majeski_2024arXiv240502418M}. Here, we proceed to discuss the implications of dominant micromirror transport, assuming that $f_\mathrm{mm}$ is not tiny, viz., $f_\mathrm{mm} \gtrsim 0.1$, as indeed observed in recent numerical simulations \citep{St-Onge2018ApJ...863L..25S, St-Onge2020JPlPh..86e9003S, Zhou_2023, Majeski_2024arXiv240502418M}. Studies of Faraday depolarization of radio emission from radio galaxies could be in principle used to constrain the volume-filling fraction of micromirrors, because depolarization increases proportionally to~$f_\mathrm{mm}$. We tested the expected Faraday depolarization arising from micromirrors within galaxy clusters in a numerical experiment using our PIC simulations (Section~\ref{sec:mm_sim}). The expected depolarization angles due to micromirror fluctuations for wavelengths observed with VLT and LOFAR are too small to constrain $f_\mathrm{mm}$. Details of the resolution element in radio telescope observations may affect this result, an issue reserved for future studies.

\section{Discussion}\label{sec3}
We have argued that CR diffusion in the ICM is determined by microscale ($l_\mathrm{mm}$) mirrors at CR energies ${\rm GeV}\lesssim E\lesssim{\rm TeV}$ and by macroscale ($l_\mathrm{c}$) turbulence at $E\gtrsim{\rm TeV}$. 
Although the micro-macrophysics transition mainly depends on these two scales, there is a degree of fine-tuning at mesoscales. This refers to the role played in~(\ref{eq:energy}) by the exponent~$\delta$, which depends on the details of the scattering mechanism and of the turbulent cascade (see Section~\ref{sec:mesoscales}). While our study is tailored to the ICM, it can be adapted to other nearly collisionless high-$\beta$ plasmas, such as the hot interstellar medium and the Milky Way halo. In what follows, we discuss what this revised picture of CR diffusion in the ICM implies for our understanding of (Fermi/radio) bubbles and other similar large-scale morphologies.

The reduction of CR diffusion caused by micromirrors may act as a transport barrier in the ICM, reducing the escape of sub-TeV CRs from their sources. For example, as radio bubbles rise through the ICM, their confinement of CRs may be dominated by micromirror confinement rather than by the conventional mechanism of magnetic draping; this depends on the details of diffusion coefficients both parallel and perpendicular to the mean magnetic field. (see, e.g., \cite{Ensslin_2003,Dursi_Pfrommer_2008,Ruszkowski_2008MNRAS.383.1359R}).~\cite{Ewart_2024arXiv240405110E} considered the CR transport in such systems, and showed that the sub-TeV CRs form a thin layer on the surface of the bubble and acquire hardened spectral-density distributions. Such sharp boundaries and hardened CR spectra, which translate into hardened photon spectra, are indeed observed in popular morphologies like the Fermi Bubble \citep{Herold2019A&A...625A.110H} and the radio bubble in the Ophiuchus galaxy cluster \citep{Giacintucci2024}. The decreased diffusion coefficients of sub-TeV CRs could also be relevant in high-$\beta$ regions within galaxies. One possible example is the Cygnus Cocoon, a Galactic PeVatron, located within a star-forming region, where the observationally constrained suppressed CR diffusion coefficients \citep{2021NatAs...5..465A, 2024SciBu..69..449L} may be explained by the additional collisionality due to micromirrors.

We have shown that micromirror diffusion significantly reduces the CR mean free path, thus, in principle, making CR coupling to the ambient motions tighter. However, in order to assess what this does to the efficiency (or otherwise) of the (re)acceleration \citep{Miniati_2015, Brunetti2011MNRAS.412..817B, Brunetti_2016MNRAS.458.2584B, Bustard_2022ApJ...941...65B, Lazarian2023ApJ...956...63L, Beduzzi_2023arXiv230603764B, Nishiwaki_2024}, one must have a somewhat more detailed picture than we currently do of the nature of the ICM turbulence \citep[i.e., of turbulence in a weakly collisional, high-beta plasma—a topic active current investigations; see][]{Arzamasskiy_2023PhRvX..13b1014A, Squire_2023arXiv230300468S, Majeski_2024arXiv240502418M} and of how the micromirror patches might be shaped and spatially distributed in this turbulence. This will require further study before the (re)acceleration question is settled.

A further example of how micromirror diffusion may impact the surrounding plasma arises from the observation that, as the scattering of sub-TeV CRs at micromirrors increases the effective CR collisionality in high-$\beta$ environments, 
the effective operation of the CR streaming instability within micromirror patches is put into doubt (see Section~\ref{sec:streaming_inst}). Note that a patchy distribution of the micromirrors may allow for the existence of regions where the streaming instability remains active -- indeed possibly more so than usually expected, as those micromirror-free regions are likely to feature negative pressure anisotropies and, therefore, reduced effective Alfv\'en speeds. Models designed to explain the thermal balance between heating and cooling of galaxy clusters based on collisionless, resonant mechanisms \citep{Guo2008MNRAS.384..251G, Jacob_2017MNRAS.467.1449J, Jacob_2017MNRAS.467.1478J}, thus become less plausible.

Finally, the micromirror scattering matters for cosmological studies of the evolution of the ICM and galaxy clusters. The suppressed CR diffusion coefficients offer a compelling justification for how CRs can be effectively ``frozen'' within the ICM as key parameters such as gas density and magnetic-field intensity evolve -- an assumption fundamental to recent models of the dynamical evolution of galaxy clusters and their surroundings \citep{Cuciti_2022Natur.609..911C, Beduzzi_2023arXiv230603764B, Boess_2023MNRAS.519..548B}. This impacts the interplay between CRs and other astrophysical processes within these massive cosmic structures \citep{Hopkins_2020MNRAS.492.3465H, Ruszkowski_2023arXiv230603141R}.

While it is well established that macroscopic dynamics can trigger microscopic phenomena, the potentially transformative impact of micromirrors on CR diffusion provides a lesson that microphysics can reciprocally affect macroscopic dynamics and observable structures across a range of astrophysical scales.

\section{Methods}\label{sec:method}
Modeling CR transport across a wide energy spectrum from GeV to EeV energies in a multiscale high-$\beta$ plasma presents methodological and computational challenges. For our numerical simulations, we choose parameters from the ICM, a high-$\beta$ plasma. Rotation-measure data indicate magnetic-field strengths ${\sim}0.1$--$1\,\mu$G averaged over a Mpc$^3$ ICM volume \citep{Govoni_2017A&A...603A.122G,Stuardi2021}, with typical field strengths of several $\mu$G in central regions \citep[e.g.,][]{Bonafede2010}. Numerical simulations support these estimates \citep[e.g.,][]{Steinwandel_2023arXiv230604692S}. We choose $B \sim 3\,\mu$G, $\delta B_\mathrm{mm}/B \sim 1/3$, the micromirror scale~$l_\mathrm{mm} \sim 100\,$npc, and turbulence correlation length~$l_{\rm c} \sim 100\,$kpc \citep{Kunz_2022hxga.book...56K}. Simulations are performed using the publicly available tool CRPropa 3.2 \citep{2022JCAP...09..035A} with additional extensions and modeling choices described below.

\subsection{Model of synthetic magnetic turbulence}\label{sec:synthetic_turbulence}
Modeling the competing micro- and macrophysical transport effects requires resolving turbulence over at least ten decades in scale. Current MHD and PIC simulations are unsuitable for this as they only allow for limited scale ranges \citep{Mertsch_2020}. Even the current best grid resolutions of more than $10^{12}$ cells resolve less than four decades of scale separation. Synthetic turbulence, on the other hand, can be generated by summing over $n_\mathrm{m}$ plane waves at an arbitrary particle position $\boldsymbol{r}$ as follows \citep{Giacalone_1999ApJ...520..204G, Tautz_2013PhPl...20b2302T}:
\begin{equation}
\begin{split}
    \delta \boldsymbol{B}(\boldsymbol{r}) &= \mathrm{Re} \left(\sum_{n=1}^{n_m}  \delta \boldsymbol{B}_n^*\,\mathrm{e}^{{\rm i}\boldsymbol{k}_n\boldsymbol{\cdot} \boldsymbol{r}}\right) \\
    &= \sqrt{2} \,\delta B \sum_{n=1}^{n_m} \boldsymbol{\xi}_n A_n\,\cos\left(k_n\, \hat{\boldsymbol{k}}_n\boldsymbol{\cdot} \boldsymbol{r} + \phi_n\right)
    \,,
\end{split}
\end{equation}
with normalized amplitudes $A_n$ determined by the assumed turbulent energy spectrum, uniformly distributed phase factors~$\phi_n~\in~[0, 2\pi]$, unit wavevectors $\hat{\boldsymbol{k}}_n$, and polarizations $\boldsymbol{\xi}_n$, satisfying~$\hat{\boldsymbol{k}}_n \boldsymbol{\cdot} \boldsymbol{\xi}_n = 0$. We employ the performance-optimized method described in \cite{Schlegel_2020}. We investigated the number of wavemodes $n_\mathrm{m}$ needed by analyzing turbulence characteristics and diffusion coefficients of CRs and found that $n_\mathrm{m} = 1024$ log-spaced wavemodes are sufficient for diffusion coefficients to converge.

We compared our key results on CR transport obtained with the above method to those obtained with an alternative method for synthetic turbulence, where we followed the approach proposed, e.g., in \cite{Giacinti_2012} and  \cite{Mertsch_2020}. In this alternative method, synthetic turbulence is precomputed and stored on many discrete nested grids at different scales, with magnetic fluctuations between scales $l_{\mathrm{min},i}$ and $l_{\mathrm{max},i}$ with individual magnetic-field strengths $\delta B_i^2 = \delta B^2 (l_{\mathrm{max},i}^{\xi-1} -l_{\mathrm{min},i}^{\xi-1})/(l_{\mathrm{max}}^{\xi-1} -l_{\mathrm{min}}^{\xi-1})$ and $l_\mathrm{max}\sim 5 l_\mathrm{c}$. We found agreement between results obtained using both methods.

Note that CRs below PeV energies diffuse on time scales smaller than the typical timescale of large-scale turbulence motions, justifying our modeling turbulence as static. CRs are ``frozen'' within the ICM as the plasma evolves, as discussed in Section~\ref{sec3}.

\subsection{CR trajectories}\label{sec:mathod_eqm}
Computing CR trajectories in magnetic fields involves solving the equation of motion for charged particles. These are then used to calculate the statistical transport characteristics of CRs. We use the Boris-push method for this task, as implemented in \cite{2022JCAP...09..035A}. This captures the dynamics of charged particles in magnetic fields while preserving key properties, such as CR energy.

\subsection{Model of small-angle scattering in magnetic micromirrors}\label{sec:mathod_mmm}
We model the effect of magnetic micromirrors as a change of propagation after a distance~$s$ by an angle $\delta \theta$ given the scattering rate $\nu_\mathrm{mm} = \delta \theta^2\, c/s$. Particles traveling the mean free path $\lambda_\mathrm{mm} = c / \nu_\mathrm{mm}$ will have lost information of their original direction. At each step $s$, chosen to be smaller than $\lambda_\mathrm{mm}$, we introduce a small deflection 
\begin{equation}
    \delta \theta = X \, \sqrt{\frac{s\,\nu_\mathrm{mm}}{c}}\, ,
\end{equation}
where the random Gaussian variable $X$ with mean 0 and standard deviation~1 represents the assumption that CRs random-walk their way through the magnetic micromirrors. 
Alternatively, micromirrors could be directly modeled in the magnetic field, which would, however, necessitate step sizes $s \lesssim l_\mathrm{mm} \ll \lambda_\mathrm{mm}$, leading to significantly longer simulation times.

\subsection{Model of CR scattering at mesoscales}\label{sec:mesoscales}
There is an ongoing debate about the dominant mechanism of CR scattering off mesoscale fluctuations, with theories including ``extrinsic'' (cascading) turbulence and ``self-excitation'' (by kinetic CR-driven instabilities) scenarios \citep[see][for recent overviews]{Kempski_2022MNRAS.514..657K, Hopkins_2022MNRAS.517.5413H}. Here, we focus on the extrinsic scenario, with the self-excitation described in Section~\ref{sec:streaming_inst}.

The CR diffusion coefficient due to resonant scattering is, by dimensional analysis,
\begin{equation}\label{eq:kappa_res_rg}
    \kappa_\mathrm{res} \sim  \frac{c\,r_\mathrm{g}}{f(r_\mathrm{g})}\,,
\end{equation}
where $f(r_\mathrm{g})$ is a dimensionless model-dependent numerical factor expressing the efficiency of the resonant scattering off turbulent magnetic structures at the scale~$l \sim r_\mathrm{g}$. We assume diffusion rather than superdiffusion, which would increase diffusion coefficients with time. This serves as a conservative estimate. Quasi-linear theory \citep{Jokipii1966} determines the factor $f(r_\mathrm{g})$ for isotropic turbulence as the fraction of the parallel turbulent power located at the gyroresonant scales $l = 2\pi/k_\parallel \sim r_\mathrm{g}$, viz., $f(r_\mathrm{g})\sim\int_{2\pi/r_\mathrm{g}}^\infty \mathrm{d}k_\parallel \,P(k_\parallel) /(B^2/8\pi) \leq 1$, where $k_\parallel$ is the parallel wavenumber and $P(k_\parallel)$ is the parallel magnetic-energy spectrum.
Assuming an undamped turbulent cascade with $P(k_\parallel) \propto k_\parallel^{-\xi}$ gives $f(r_\mathrm{g}) \sim (r_\mathrm{g}/l_\mathrm{c})^{\xi-1}$, where $l_\mathrm{c}$ is the energy-containing scale. Therefore,
\begin{equation}\label{eq:kappa_res_app}
    \kappa_\mathrm{res} \sim c\,l_\mathrm{c} \left(  \frac{r_\mathrm{g}}{l_\mathrm{c}} \right)^{2-\xi} \propto E^{2-\xi}l_\mathrm{c}^{\xi-1} = E^{\delta}l_\mathrm{c}^{-\delta+1}\,,
\end{equation}
where $\delta = 2-\xi$ is defined for convenience.
In Section~\ref{sub:transition}, we confirmed this scaling via numerical simulations with unprecedented spatial resolution for a synthetic turbulence composed of plane waves.

An important qualitative result is that the cases $\delta = 2$ ($\xi = 0$) 
and $\delta = 1$ ($\xi = 1$) apply only to $r_\mathrm{g} \gg l_\mathrm{c}$ and $r_\mathrm{g} \sim l_\mathrm{c}$, respectively. While there is no realistic turbulence model with $\xi = 0$, this case formally corresponds to the small-angle scattering limit of CRs with $r_\mathrm{g} \gg l_\mathrm{c}$:
equation~(\ref{eq:kappa_res}) then yields $\kappa \sim c\,r_\mathrm{g}^2/l_\mathrm{c}$, which is identical to~(\ref{eq:kappa_mm}) if one replaces $l_\mathrm{mm}$ and $\delta B_\mathrm{mm}$ with $l_\mathrm{c}$ and $B$, respectively. This is the standard theory for the high-energy regime referred to in the Introduction. 

In fact, only scalings with weaker energy dependence for~$r_\mathrm{g} \ll l_\mathrm{c}$ are typically considered. In this limit, three popular choices for this scaling have appeared in the literature: (i) $\delta = 1/3$, corresponding to isotropic turbulence with a \cite{Kolmogorov_1941DoSSR..30..301K} spectrum ($\xi = 5/3$); (ii) $\delta = 1/2$, corresponding to $\xi = 3/2$, which in the past was associated with the theory by \cite{Iroshnikov_1964SvA.....7..566I} and \cite{Kraichnan_1965PhFl....8.1385K} for weak, isotropic Alfv\'{e}nic magnetohydrodynamic (MHD) turbulence (now known not to exist); and (iii) $\delta = 0$, corresponding to the $\xi = 2$ Goldreich--Sridhar parallel spectrum of critically balanced Alfv\'{e}nic turbulence
(\cite{GS_1995ApJ...438..763G}; see \cite{Schekochihin_2022JPlPh..88e1501S} for a review) by adhering to~(\ref{eq:kappa_res}) \cite[note that this is not observed: see, e.g.,][]{Fornieri_2021MNRAS.502.5821F}; nevertheless, this spectrum provides a simple way to estimate the decreased CR-scattering efficiency expected for the anisotropic turbulent cascade). Historically, Alfv\'enic turbulence was favored until it was realized that scale-dependent anisotropy \citep{Chandran_2000PhRvL..85.4656C, Lithwick_2001ApJ...562..279L, Cho_2000ApJ...539..273C}, damping \citep{Skilling_1975MNRAS.173..255S, Hopkins_2022MNRAS.517.5413H}, and intermittency \citep{Shukurov_2017ApJ, Perri_2019SoPh..294...34P} might lead to inefficient gyroresonant scattering. A putative $\xi=3/2$ cascade of fast MHD modes, if isotropic and robust against steepening \citep{Kadomtsev_1973SPhD...18..115K, Kempski_2022MNRAS.514..657K} and various damping mechanisms, may help by generating fluctuations with large enough frequencies and amplitudes to scatter CRs efficiently \citep{Yan_2002PhRvL..89B1102Y, Cho_2002PhRvL..88x5001C}.
More recently, the exponents $\delta = 1/3$ and $\delta = 1/2$ were ascribed to CR scattering in intermittent distributions of sharp magnetic-field bends in Goldreich--Sridhar turbulence \citep{Lemoine_2023arXiv230403023L} and in MHD turbulent dynamo \citep{Kempski_2023arXiv230412335K}. 
Scaling exponents in the range $0.3 \lesssim \delta \lesssim 0.5$ are in broad agreement with constraints from Galactic observations \citep[see][for a review]
{Tjus_2020PhR...872....1B}.

An additional process that may contribute to the diffusion of low-energy CRs arises from CRs following diffusing magnetic-field lines. The Alfvénic scale $l_\mathrm{A}\gtrsim 1\,$kpc \citep{Hu2020ApJ...901..162H} approximates the mean free path of CRs following these field lines \citep{Brunetti_2007MNRAS.378..245B}. The associated CR diffusion coefficient $\kappa_\mathrm{flrw} \sim c\,l_\mathrm{A} \gtrsim 10^{32}\,$cm$^2$~s$^{-1}$ does not fall significantly below $\kappa_\mathrm{res}$ for $r_\mathrm{g} \ll l_\mathrm{c}$. Given this estimate, we neglect this transport process in our simulation setup, but indicate the value of the diffusion coefficient corresponding to it as an upper boundary in Figure~\ref{fig:result}.

\subsection{CR streaming instability}\label{sec:streaming_inst}
Let us now explain why the streaming instability can be ignored in our multiscale model of CR transport within micromirror patches. The self-confinement of CRs due to the streaming instability is believed to play an important role in the Galaxy \citep{Blasi_2012PhRvL.109f1101B} and in galaxy clusters \citep{Jacob_2017MNRAS.467.1449J}.
In this picture, the streaming instability generates fluctuations of the magnetic field, which in turn can scatter CRs. It is believed that this mechanism may take over at lower CR energies, with details depending on the instability's growth rate at wavenumber $k$. At gyroscale, \citep{Kulsrud_1969ApJ...156..445K}:
\begin{equation}\label{eq:growth_rate}
\begin{split}
    \gamma_\mathrm{SI} &\sim \Omega_i \frac{n_\mathrm{CR}({>}E)}{n_i} \left(\frac{v_\mathrm{st}}{v_\mathrm{A}} - 1\right) \\
    &\sim 10^{-14}\,\left(\frac{B}{3\,\mu\mathrm{G}}\right) \left(\frac{E}{\mathrm{TeV}}\right)^{-1.6}\,\mathrm{s}^{-1}\,,
\end{split}
\end{equation}
where ${n_\mathrm{CR}({>}E)}$ is the density of CRs with energies above energy $E$ corresponding to the resonance condition that can interact resonantly with waves with wavenumber $k$, $n_i$ is the ambient ion density, $v_\mathrm{A}$ is the Alfvén speed, and $v_\mathrm{st}$ is the streaming speed, believed to be on the order of $v_\mathrm{A}$ in saturation for the $\sim \,$GeV CRs \citep{Jacob_2017MNRAS.467.1449J, Krumholz_2020MNRAS.493.2817K}. In the second estimate in~(\ref{eq:growth_rate}), we employed the common assumptions \cite[see, e.g.,][and references therein]{Zweibel_2013PhPl...20e5501Z, Kempski_2020MNRAS.493.5323K} that $(v_\mathrm{st}/v_\mathrm{A}-1) \sim 1$ and $n_\mathrm{CR}({>}E)/n_i \sim 10^{-7} (E/\mathrm{GeV})^{1-\alpha}$ in galaxy clusters, with $\alpha \approx 2.6$.

Scattering of sub-TeV CRs at micromirrors increases the effective CR collisionality in high-$\beta$ environments. A comparison of the gyroscale growth rate~(\ref{eq:growth_rate}) with the scattering rate at micromirrors~(\ref{eq:nu_mm}) gives
\begin{equation}\label{eq:streaming_growth_suppressed}
    \frac{\gamma_\mathrm{SI}}{\nu_\mathrm{mm}} \sim 10^{-5} \,   Z^{-2}\left(\frac{T}{5\,\mathrm{keV}}\right)^{-1/2} \left(\frac{\delta B_\mathrm{mm}/B}{1/3} \right)^{-2} \left(\frac{E}{\mathrm{TeV}}\right)^{0.4}\,.
\end{equation}
With such a large effective collisionality isotropizing and homogenizing CRs, it is doubtful that this gyroscale, resonant instability can operate. 

Another way to gauge the importance of the streaming instability is to imagine that it is not suppressed and then check for self-consistency. In particular, for self-confined CRs, the CR-density scale height $H$ is set by the properties of the ambient thermal gas and is of order the thermal-gas-density scale height~$H_\rho$. If scattering by micromirrors is present with diffusion coefficient $\kappa_{\rm mm}$, the associated diffusive flux is smaller than the minimum flux required for the streaming instability to operate if $\kappa_{\rm mm} /H \lesssim  v_{\rm A}$. This corresponds to scattering by micromirrors suppressing the anisotropy in the CR distribution function to levels below ${\sim }v_{\rm A}/c$, which is the threshold anisotropy for the streaming instability to operate in the first place. Thus, because $\kappa_{\rm mm} /H  \lesssim v_{\rm A}$ for sub-TeV CRs, where $H \sim H_\rho \gtrsim 10\,$kpc \citep{Fabian2006MNRAS.366..417F, Lyskova2023MNRAS.525..898L}, CRs may not self-confine in a self-consistent manner.

\subsection{Model of CR streaming}\label{sec:streaming}

Let us imagine that the instability is not suppressed, despite the arguments made in Section~{\ref{sec:streaming_inst}}, and estimate \cite{Brunetti_2007MNRAS.378..245B}:
\begin{equation}\label{eq:kappa_st}
    \kappa_\mathrm{st} \sim l_\mathrm{A} v_\mathrm{st} \gtrsim l_\mathrm{A} v_\mathrm{A} \gtrsim 3 \cdot 10^{28}\, \left(\frac{l_\mathrm{A}}{1\,\mathrm{kpc}} \right)\left(\frac{v_\mathrm{A}}{100\,\mathrm{km/s}} \right)\,\mathrm{cm}^2~\mathrm{s}^{-1}\,,
\end{equation}
where we have used the Alfvén speed $v_\mathrm{A} \sim 100\,\mathrm{km/s}$ as the lower limit of the streaming speed. We also assumed that the magnetic-field lines stochastically tangled on the scale $l_\mathrm{A}$. In super-Alfvénic turbulence, this marks the transition towards fully magnetohydrodynamic turbulence \citep[e.g.,][]{Brunetti_2007MNRAS.378..245B}, as it is the scale at which the turbulent velocity matches the Alfvén speed. This ``Alfvén scale'' $l_\mathrm{A}$, which is $\gtrsim 1\,$kpc under typical ICM conditions  \citep{Hu2020ApJ...901..162H}.

Comparing this CR diffusivity with the one caused by micromirrors~(\ref{eq:kappa}) gives 
\begin{equation}\label{eq:k_mm_k_st}
\begin{split}
    \frac{\kappa_\mathrm{mm}}{\kappa_\mathrm{st}} \lesssim Z^{-2}&\left(\frac{E}{100\,\mathrm{GeV}} \right)^2  
    \left(\frac{l_\mathrm{A}}{1\,\mathrm{kpc}} \right)^{-1}\left(\frac{T}{5\,\mathrm{keV}} \right)^{-1/2}\\&\left(\frac{B}{3\,\mu\mathrm{G}} \right)^{-1} \left(\frac{\delta B_\mathrm{mm}/B}{1/3} \right)^{-2}
    \left(\frac{v_\mathrm{A}}{100\,\mathrm{km/s}} \right)^{-1}\,,
\end{split}
\end{equation}
demonstrating that $\kappa_\mathrm{mm}$ dominates CR transport in the ICM up to almost TeV energies. This comparison, together with the arguments for the suppression of the streaming instability (Section~{\ref{sec:streaming_inst}}), justifies neglecting CR streaming in our numerical experiments. Nevertheless, we show this estimate in Figure~\ref{fig:result}.

\subsection{Computation of CR diffusion coefficient}\label{sec:RNN}
We use the CR propagation software CRPropa 3.2 \citep{2022JCAP...09..035A} and extend the framework with our custom modules for the generalized nested turbulence, different turbulence geometries, and micromirror scattering. We choose sufficiently small step sizes~$s_\mathrm{step} \sim \min \{\lambda_\mathrm{mm}/10^3, \lambda_\mathrm{res}/10^3, \lambda_\mathrm{he}/10^3\}$, with mean free path $\lambda_i \sim \kappa_i / c$ to resolve the small-angle scattering at micromirrors, the resonant scattering in the extrinsic turbulent cascade, and the small-angle scattering in the high-energy limit, respectively. The option $\lambda_\mathrm{he}/10^3$ is only included for CR energies above $10\,$EeV, as the high-energy limit is not valid below that energy. 
We compute sufficiently long CR trajectories $d \sim \min \{10^3\lambda_\mathrm{mm}, 10^3\lambda_\mathrm{res}, l_\mathrm{max}\}$ (min is used here to save computation resources as only the more efficient scattering process needs to be resolved) for $E \lesssim 10\,$EeV and $d \sim \max \{10^3\lambda_\mathrm{he}, l_\mathrm{max}\}$ otherwise. The time-dependent diffusion coefficient $\kappa(t)$ for CRs performing a correlated random walk is \citep{chen_renshaw_1992}
\begin{equation}
    \kappa(t) = \frac{ \left \langle \Delta r^2 \right \rangle}{2t} \left(1+\frac{2\left \langle \cos \Theta \right \rangle}{1 - \left \langle \cos \Theta \right \rangle} \right) \stackrel{t \gg \lambda/c}{\approx} \frac{ \left \langle \Delta r^2 \right \rangle}{2t}\,,
\end{equation}
with $r$ representing the CR spatial displacement and the operation $\left \langle ... \right \rangle$ averaging over CRs. The approximation for $t \gtrsim \lambda/c$ stems from the convergence to central-limit behavior, assuming that the deflection angles $\Theta$ are uniformly distributed after CRs travel a distance equivalent to their mean free path $\lambda$. We compute the steady-state diffusion coefficient $\kappa$ by averaging $\kappa(t)$ for $10^3$ CRs. When using synthetic turbulence, we average this quantity over ten different realizations.

To compute the CR diffusion coefficients efficiently, we have developed a recurrent neural network (RNN) that comprises a long short-term memory (LSTM) layer followed by a fully connected layer, linking the output of the LSTM to our output size, which is the predicted diffusion coefficient. This architecture allows us to capture the temporal dependence in our data, making the network well-suited to use the CR trajectories as input for the model. For training data, we used trajectories generated by Monte Carlo simulations using the method described in Section~\ref{sec:mathod_two_phase_medium}, with various different effective diffusion coefficients. These effective diffusion coefficients served as labels for the training process. We chose low statistics of only 400 CRs for the 1D cases to illustrate the superiority of our neural network (trained on only 1600 CR trajectories in less than a minute on a conventional CPU) over the classical computation of the running diffusion coefficient. We tested the convergence of the latter method to the theoretical expectation in the limit of large times and CR numbers and found good agreement. We demonstrate the capabilities of the model in Figure~\ref{fig:result_fmm}. The good performance of the network is primarily attributable to its capacity to learn efficiently that the diffusion coefficients can be predicted accurately by the frequency of small-angle scattering, which becomes evident in a relatively small number of steps. Additionally, the ability to distinguish signal (deflection caused by fluctuations) from noise (constant gyration) further contributes to the network's superior performance. These abilities indicate that the network could also be employed as a robust framework for assessing the micromirror filling fraction $f_\mathrm{mm}$ in MHD turbulence by propagating charged particles through the magnetic field. Reliable computation of the mean-squared diffusion coefficient requires many trajectories, necessary to deduce $f_\mathrm{mm}$. Our use of a neural network demonstrates a concept for efficient trajectory classification and the computation of transport characteristics in astrophysical systems.

\subsection{Micromirror field from PIC simulations}\label{sec:mm_sim}
The micromirror field shown in Figure~\ref{fig:mirrorfield} was self-consistently generated using the hybrid-kinetic code \textsc{Pegasus}\texttt{++}~, which models the collisionless ions using a PIC method and the electrons as an isotropic, isothermal fluid. The code can simulate a plasma’s expansion or contraction using a coordinate transform method \cite[see][for further details]{Bott_2021ApJ...922L..35B}, which, via double-adiabatic conservation laws, produces an ion pressure anisotropy that becomes mirror unstable. 

In our simulation, we initialized a uniformly magnetized plasma ($\boldsymbol{B}_0 = B_0 \hat{\boldsymbol{x}}$) with a Maxwellian population (3000 ion macro-particles per cell) on a cubic domain. Its size is $L_0^3 = (76.8 r_{\mathrm{g},i0})^{3}$, where $r_{\mathrm{g},i0}$ is the initial gyroradius of thermal ions, and the grid resolution is $\Delta x = 0.1 r_{\mathrm{g},i0}$, $\Delta y = \Delta z = 0.3 r_{\mathrm{g},i0}$. The initial ion plasma beta is $\beta_{i0} = 50$. The scale $L_{\perp}$ of the plasma in the direction perpendicular to the background field then evolves as $L_{\perp} = L_{0} (1+t/\tau_{\rm crt})^{-2}$, with a contraction timescale of $\tau_{\rm crt} = 5 \times 10^{3} \Omega_{i0}^{-1}$, while the parallel scale remains fixed. This gives rise to an ion pressure anisotropy $\Delta_i \equiv T_{\perp, i}/T_{\parallel, i}-1 = (1+t/\tau_{\rm crt})^{2} - 1$ that increases with time. The mirror instability is triggered at $t \approx \tau_{\rm crt}/2 \beta_{i0} \simeq 0.01 \tau_{\rm crt}$, and then back-reacts at $t \approx 0.1  \tau_{\rm crt}$, entering the secular (i.e., power-law) phase of growth \citep{Kunz_2014PhRvL.112t5003K, Rincon_2015MNRAS.447L..45R}. The simulation is then run until $t_{\rm end} \approx 0.25  \tau_{\rm crt}$. The snapshots of the field shown in Figure~\ref{fig:mirrorfield} were taken at $t \approx 0.15 \tau_{\rm crt}$ and $t = t_{\rm end}$, respectively.  

Like in all PIC simulations, the finite number of macroparticles in \textsc{Pegasus}\texttt{++} leads to grid-scale noise in the electromagnetic fields. To diagnose the influence of this noise on our calculation of CR propagation, we performed an experiment in which we removed the micromirrors from our \textsc{Pegasus}\texttt{++} simulation using a Fourier filter, and integrated CRs through the residual magnetic field. While the resulting diffusion coefficients show that PIC noise also leads to diffusive CR transport, their values are much larger than the ones associated with the micromirrors and so can be safely ignored (see Figure~\ref{fig:mirror_kappa}).

\subsection{Turbulence from MHD simulations}\label{sec:mhd_sim}
At macroscales, we computed CR diffusion coefficients in forced incompressible MHD turbulence from the John Hopkins Turbulence Databases \citep{Eyink_2013Natur.497..466E, JHTDB} to validate the consistency of our numerical approach that relies on synthetic turbulence. The MHD turbulence was generated in a direct numerical simulation of the incompressible MHD system of equations without guide fluid using $1024^3$ nodes, employing a pseudo-spectral method, with energy input from a Taylor-Green flow stirring force. CR diffusion coefficients in this field are shown in Figure~\ref{fig:result}.

\subsection{Model of a static two-phase inhomogeneous medium}\label{sec:mathod_two_phase_medium}
We model the CR transport through a two-phase inhomogeneous medium in 1D and in 3D (see a visualization in Figure~\ref{fig:visualization}). 
In 3D, we modified the computationally intensive numerical experiment described in the previous subsections as follows: instead of imposing an additional effective scattering rate $\nu_\mathrm{mm}$ on all CRs propagating through our turbulent field at all times and places, we now turn this scattering on only if the CR is seeing a magnetic field above a certain threshold $B_>$. This models qualitatively the fact that micromirrors are likely to appear in regions of more vigorous magnetic-field amplification. By varying $B_>$ between $0$ and $\infty$, we effectively vary the micromirror volume-filling fraction between $1$ and $0$, respectively. We assume the most intuitive interpretation that this volume-filling fraction is the effective micromirror fraction $f_\mathrm{mm}$ in~(\ref{eq:nu_eff_f}).
As our estimate of the effective diffusion coefficient works in all dimensionalities, we also model the CR transport in a two-phase inhomogeneous medium in 1D. This allows us to compute diffusion coefficients effectively for many different values of $f_\mathrm{mm}$. In doing so, we employ a simplified Monte Carlo model with $\nu(x) = \nu_\mathrm{mm}$ for $x$ mod $1\leq f_\mathrm{mm}$ and $\nu(x) = \nu_\mathrm{res}$ otherwise.

We now justify modeling micromirror patches experienced by diffusing CRs as static, based on the assumption of the short residence time of CRs in patches
\begin{equation}
    \Delta t_\mathrm{p} \sim \frac{l_\mathrm{p}}{c}\,,
\end{equation}
where $l_\mathrm{p}$ is the characteristic size of the patch.
The residence time of CRs in a patch can be understood intuitively: when CRs penetrate a patch, they do so ballistically up to their mean free path~$\lambda_\mathrm{mm} \sim \kappa_\mathrm{mm}/c$ for a time~$\Delta t_\mathrm{bal} \sim \lambda_\mathrm{mm}/c$, followed by isotropic diffusion over a time~$\Delta t_\mathrm{diff}$. To get a rough estimate for~$\Delta t_\mathrm{diff}$, we can consider the simplified 1D case, where CRs exit either at the point of entry or at the point on the opposite end of the patch, with probabilities~$p_1 \sim (l_\mathrm{p}-\lambda_\mathrm{mm})/l_\mathrm{p}$ and~$p_2 \sim \lambda_\mathrm{mm}/l_\mathrm{p}$, respectively. The corresponding times needed to exit the patch via diffusive transport are~$\Delta t_1 \sim \lambda_\mathrm{mm}^2/\kappa_\mathrm{mm}$ and~$\Delta t_2 \sim (l_\mathrm{p}-\lambda_\mathrm{mm})^2/\kappa_\mathrm{mm}$. Note that in the limit $\lambda_\mathrm{mm} \ll l_\mathrm{p}$, $\Delta t_2\sim l_\mathrm{p}^2/c\lambda_\mathrm{mm}$ can be larger than $\Delta t_\mathrm{lifetime}$, meaning that patches cannot be treated as being static anymore. However, this only affects a small fraction $p_2 \sim \lambda_\mathrm{mm}/l_\mathrm{p} \ll 1$ of CRs.
The mean time duration of the diffusive transport is then given by
\begin{equation}
\begin{split}
    \Delta t_\mathrm{diff} &\sim p_1 \Delta t_1 + p_2 \Delta t_2 \\
    &\sim \frac{l_\mathrm{p}-\lambda_\mathrm{mm}}{l_\mathrm{p}}
    \frac{\lambda_\mathrm{mm}^2}{c\lambda_\mathrm{mm}} + \frac{\lambda_\mathrm{mm}}{l_\mathrm{p}}
    \frac{(l_\mathrm{p}-\lambda_\mathrm{mm})^2}{c\lambda_\mathrm{mm}} = \frac{l_\mathrm{p}-\lambda_\mathrm{mm}}{c}\,,
\end{split}
\end{equation}
resulting in~$\Delta t_\mathrm{p} \sim \Delta t_\mathrm{bal} + \Delta t_\mathrm{diff} \sim l_\mathrm{p}/c$. \cite{Kotera_2008PhRvD..77l3003K} obtained this result by a more general method of solving the time-dependent diffusion equation. It means that CRs with small diffusion coefficients, corresponding to short mean free paths, will exit the patch dominantly near their point of entry. In contrast, CRs with large diffusion coefficients traverse the patch (quasi)ballistically in time~$\Delta t_\mathrm{p} \sim \Delta t_\mathrm{bal} \sim l_\mathrm{p}/c$. As $\Delta t_\mathrm{diff} \ll \Delta t_\mathrm{lifetime}$, micromirrors can be treated as being static for most CRs.\\\\
\textbf{Correspondence and requests for materials} should be addressed to
P.~Reichherzer.

\section*{Data availability}
The primary data, including diffusion coefficients, energies, and simulation parameters, is available at \cite{ZENODO}. The magnetic field data will be made available at reasonable request to the corresponding author.

\section*{Code availability}
CR simulations were performed with CRPropa3, specifically with the version \url{https://github.com/reichherzerp/CRPropa3}.

\section*{Acknowledgments}
PR thanks J.~Becker~Tjus, F.~Effenberger, H.~Fichtner, R.~Grauer, J.~L\"{u}bke, and L.~Schlegel for valuable discussions on the usage of synthetic turbulence. AAS thanks E.~Churazov and S.~Komarov for useful discussions of the ICM and CR physics. The authors thank P.~Oh, E.~Quataert, L.~Silva, L.~Turica, D.~Uzdensky, and especially W.~Xu for insightful discussions of CR transport in the ICM. We also acknowledge the generous hospitality of the
Wolfgang Pauli Institute, Vienna, where these ideas were discussed during the 14th Plasma Kinetics Working Meeting (2023).

The work of PR was initially funded through a Gateway Fellowship and subsequently through the Walter Benjamin Fellowship by the Deutsche Forschungsgemeinschaft (DFG, German
Research Foundation) --- grant number 518672034. AFAB's work was supported by a UKRI Future Leaders Fellowship (grant number MR/W006723/1). RJE’s work was supported by a UK EPSRC studentship. MWK's work was supported in part by NSF CAREER Award No.~1944972. The work of AAS and GG was supported in part by a grant from STFC (ST/W000903/1). The work of AAS was also supported by a grant from EPSRC (EP/R034737/1) and by the Simons Foundation via a Simons Investigator Award. PK was supported by the Lyman Spitzer, Jr. Fellowship at Princeton University. UKRI is a Plan S funder, so for the purpose of Open Access, the author has applied a CC BY public copyright license to any Author Accepted Manuscript version arising from this submission.

\section*{Author contributions}
AFAB, RJE, PR, and AAS conceptualized the study and developed the theoretical framework. AFAB and MWK conducted the micromirror-field simulations, while PR performed the cosmic-ray simulations, neural-network implementation and training, data analysis, and visualization. PR wrote the initial draft and led the process of revision, with contributions to writing from AFAB, RJE, MWK, PK, and AAS. All authors provided feedback, participated in data interpretation, and were involved in finalizing the manuscript. The project was supervised by AFAB, GG, and AAS.

\newpage


\begin{thebibliography}{100}
\expandafter\ifx\csname url\endcsname\relax
  \def\url#1{\burl{#1}}\fi
\expandafter\ifx\csname urlprefix\endcsname\relax\def\urlprefix{URL }\fi
\providecommand{\bibinfo}[2]{#2}
\providecommand{\eprint}[2][]{\url{#2}}
\providecommand{\doi}[1]{\url{https://doi.org/#1}}
\bibcommenthead

\bibitem{Schekochihin_2005ApJ...629..139S}
\bibinfo{author}{{Schekochihin}, A.~A.}, \bibinfo{author}{{Cowley}, S.~C.}, \bibinfo{author}{{Kulsrud}, R.~M.}, \bibinfo{author}{{Hammett}, G.~W.} \& \bibinfo{author}{{Sharma}, P.}
\newblock \bibinfo{title}{{Plasma instabilities and magnetic field growth in clusters of galaxies}}.
\newblock \emph{\bibinfo{journal}{\apj}} \textbf{\bibinfo{volume}{629}}, \bibinfo{pages}{139--142} (\bibinfo{year}{2005}).

\bibitem{Kunz_2022hxga.book...56K}
\bibinfo{author}{{Kunz}, M.~W.}, \bibinfo{author}{{Jones}, T.~W.} \& \bibinfo{author}{{Zhuravleva}, I.}
\newblock \bibinfo{title}{{Plasma physics of the intracluster medium}}.
\newblock \emph{\bibinfo{journal}{Springer Nat. Singapore}}  (\bibinfo{year}{2022}).

\bibitem{Cesarsky_1973ApJ...185..153C}
\bibinfo{author}{{Cesarsky}, C.~J.} \& \bibinfo{author}{{Kulsrud}, R.~M.}
\newblock \bibinfo{title}{{Role of hydromagnetic waves in cosmic-ray confinement in the disk. Theory of behavior in general wave spectra}}.
\newblock \emph{\bibinfo{journal}{\apj}} \textbf{\bibinfo{volume}{185}}, \bibinfo{pages}{153--166} (\bibinfo{year}{1973}).

\bibitem{Reichherzer_2020MNRAS.498.5051R}
\bibinfo{author}{{Reichherzer}, P.}, \bibinfo{author}{{Becker Tjus}, J.}, \bibinfo{author}{{Zweibel}, E.~G.}, \bibinfo{author}{{Merten}, L.} \& \bibinfo{author}{{Pueschel}, M.~J.}
\newblock \bibinfo{title}{{Turbulence-level dependence of cosmic ray parallel diffusion}}.
\newblock \emph{\bibinfo{journal}{\mnras}} \textbf{\bibinfo{volume}{498}}, \bibinfo{pages}{5051--5064} (\bibinfo{year}{2020}).

\bibitem{Lzarian_2021ApJ...923...53L}
\bibinfo{author}{{Lazarian}, A.} \& \bibinfo{author}{{Xu}, S.}
\newblock \bibinfo{title}{{Diffusion of cosmic rays in MHD turbulence with magnetic mirrors}}.
\newblock \emph{\bibinfo{journal}{\apj}} \textbf{\bibinfo{volume}{923}}, \bibinfo{pages}{53} (\bibinfo{year}{2021}).

\bibitem{Kunz_2014PhRvL.112t5003K}
\bibinfo{author}{{Kunz}, M.~W.}, \bibinfo{author}{{Schekochihin}, A.~A.} \& \bibinfo{author}{{Stone}, J.~M.}
\newblock \bibinfo{title}{{Firehose and mirror instabilities in a collisionless shearing plasma}}.
\newblock \emph{\bibinfo{journal}{\prl}} \textbf{\bibinfo{volume}{112}}, \bibinfo{pages}{205003} (\bibinfo{year}{2014}).

\bibitem{Bott_2023}
\bibinfo{author}{{Bott}, A. F.~A.}, \bibinfo{author}{{Cowley}, S.~C.} \& \bibinfo{author}{{Schekochihin}, A.~A.}
\newblock \bibinfo{title}{{Kinetic stability of Chapman-Enskog plasmas}}.
\newblock \emph{\bibinfo{journal}{\jpp}} \textbf{\bibinfo{volume}{90}}, \bibinfo{pages}{975900207} (\bibinfo{year}{2024}).

\bibitem{Aloisio_2004ApJ...612..900A}
\bibinfo{author}{{Aloisio}, R.} \& \bibinfo{author}{{Berezinsky}, V.}
\newblock \bibinfo{title}{{Diffusive propagation of ultra-high-energy cosmic rays and the propagation theorem}}.
\newblock \emph{\bibinfo{journal}{\apj}} \textbf{\bibinfo{volume}{612}}, \bibinfo{pages}{900--913} (\bibinfo{year}{2004}).

\bibitem{Kotera_2008PhRvD..77b3005K}
\bibinfo{author}{{Kotera}, K.} \& \bibinfo{author}{{Lemoine}, M.}
\newblock \bibinfo{title}{{Inhomogeneous extragalactic magnetic fields and the second knee in the cosmic ray spectrum}}.
\newblock \emph{\bibinfo{journal}{\prd}} \textbf{\bibinfo{volume}{77}}, \bibinfo{pages}{023005} (\bibinfo{year}{2008}).

\bibitem{Subedi_2017ApJ}
\bibinfo{author}{{Subedi}, P.} \emph{et~al.}
\newblock \bibinfo{title}{{Charged particle diffusion in isotropic random magnetic fields}}.
\newblock \emph{\bibinfo{journal}{\apj}} \textbf{\bibinfo{volume}{837}}, \bibinfo{pages}{140} (\bibinfo{year}{2017}).

\bibitem{Reichherzer_2022SNAS....4...15R}
\bibinfo{author}{{Reichherzer}, P.} \emph{et~al.}
\newblock \bibinfo{title}{{Regimes of cosmic-ray diffusion in Galactic turbulence}}.
\newblock \emph{\bibinfo{journal}{SN Appl. Sci.}} \textbf{\bibinfo{volume}{4}}, \bibinfo{pages}{15} (\bibinfo{year}{2022}).

\bibitem{Chen_2020ApJ}
\bibinfo{author}{{Chen}, L.~E.} \emph{et~al.}
\newblock \bibinfo{title}{{Transport of high-energy charged particles through spatially intermittent turbulent magnetic fields}}.
\newblock \emph{\bibinfo{journal}{\apj}} \textbf{\bibinfo{volume}{892}}, \bibinfo{pages}{114} (\bibinfo{year}{2020}).

\bibitem{Condorelli_2023impact}
\bibinfo{author}{{Condorelli}, A.}, \bibinfo{author}{{Biteau}, J.} \& \bibinfo{author}{{Adam}, R.}
\newblock \bibinfo{title}{{Impact of galaxy clusters on the propagation of ultrahigh-energy cosmic rays}}.
\newblock \emph{\bibinfo{journal}{\apj}} \textbf{\bibinfo{volume}{957}}, \bibinfo{pages}{80} (\bibinfo{year}{2023}).

\bibitem{Jokipii1966}
\bibinfo{author}{{Jokipii}, J.~R.}
\newblock \bibinfo{title}{{Cosmic-ray propagation. I. Charged particles in a random magnetic field}}.
\newblock \emph{\bibinfo{journal}{\apj}} \textbf{\bibinfo{volume}{146}}, \bibinfo{pages}{480} (\bibinfo{year}{1966}).

\bibitem{Giacalone_1999ApJ...520..204G}
\bibinfo{author}{{Giacalone}, J.} \& \bibinfo{author}{{Jokipii}, J.~R.}
\newblock \bibinfo{title}{{The transport of cosmic rays across a turbulent magnetic field}}.
\newblock \emph{\bibinfo{journal}{\apj}} \textbf{\bibinfo{volume}{520}}, \bibinfo{pages}{204--214} (\bibinfo{year}{1999}).

\bibitem{Lemoine_2023arXiv230403023L}
\bibinfo{author}{{Lemoine}, M.}
\newblock \bibinfo{title}{{Particle transport through localized interactions with sharp magnetic field bends in MHD turbulence}}.
\newblock \emph{\bibinfo{journal}{\jpp}} \textbf{\bibinfo{volume}{89}}, \bibinfo{pages}{175890501} (\bibinfo{year}{2023}).

\bibitem{Kempski_2023arXiv230412335K}
\bibinfo{author}{{Kempski}, P.} \emph{et~al.}
\newblock \bibinfo{title}{{Cosmic ray transport in large-amplitude turbulence with small-scale field reversals}}.
\newblock \emph{\bibinfo{journal}{\mnras}} \textbf{\bibinfo{volume}{525}}, \bibinfo{pages}{4985--4998} (\bibinfo{year}{2023}).

\bibitem{Amato_2018AdSpR..62.2731A}
\bibinfo{author}{{Amato}, E.} \& \bibinfo{author}{{Blasi}, P.}
\newblock \bibinfo{title}{{Cosmic ray transport in the Galaxy: a review}}.
\newblock \emph{\bibinfo{journal}{\asr}} \textbf{\bibinfo{volume}{62}}, \bibinfo{pages}{2731--2749} (\bibinfo{year}{2018}).

\bibitem{Luque_2023A&A...672A..58D}
\bibinfo{author}{{De La Torre Luque}, P.} \emph{et~al.}
\newblock \bibinfo{title}{{Galactic diffuse gamma rays meet the PeV frontier}}.
\newblock \emph{\bibinfo{journal}{\aap}} \textbf{\bibinfo{volume}{672}}, \bibinfo{pages}{A58} (\bibinfo{year}{2023}).

\bibitem{Cao_2023arXiv230505372C}
\bibinfo{author}{{Cao}, Z.} \emph{et~al.}
\newblock \bibinfo{title}{{Measurement of ultra-high-energy diffuse gamma-ray emission of the galactic plane from 10 TeV to 1 PeV with LHAASO-KM2A}}.
\newblock \emph{\bibinfo{journal}{\prl}} \textbf{\bibinfo{volume}{131}}, \bibinfo{pages}{151001} (\bibinfo{year}{2023}).

\bibitem{Zhang_2023arXiv230506948Z}
\bibinfo{author}{{Zhang}, R.}, \bibinfo{author}{{Huang}, X.}, \bibinfo{author}{{Xu}, Z.-H.}, \bibinfo{author}{{Zhao}, S.} \& \bibinfo{author}{{Yuan}, Q.}
\newblock \bibinfo{title}{{Galactic diffuse {\ensuremath{\gamma}}-ray emission from GeV to PeV energies in light of up-to-date cosmic-ray measurements}}.
\newblock \emph{\bibinfo{journal}{\apj}} \textbf{\bibinfo{volume}{957}}, \bibinfo{pages}{43} (\bibinfo{year}{2023}).

\bibitem{Kulsrud_1969ApJ...156..445K}
\bibinfo{author}{{Kulsrud}, R.} \& \bibinfo{author}{{Pearce}, W.~P.}
\newblock \bibinfo{title}{{The effect of wave-particle interactions on the propagation of cosmic rays}}.
\newblock \emph{\bibinfo{journal}{\apj}} \textbf{\bibinfo{volume}{156}}, \bibinfo{pages}{445} (\bibinfo{year}{1969}).

\bibitem{Skilling_1971ApJ...170..265S}
\bibinfo{author}{{Skilling}, J.}
\newblock \bibinfo{title}{{Cosmic rays in the Galaxy: convection or diffusion?}}
\newblock \emph{\bibinfo{journal}{\apj}} \textbf{\bibinfo{volume}{170}}, \bibinfo{pages}{265} (\bibinfo{year}{1971}).

\bibitem{Blasi_2012PhRvL.109f1101B}
\bibinfo{author}{{Blasi}, P.}, \bibinfo{author}{{Amato}, E.} \& \bibinfo{author}{{Serpico}, P.~D.}
\newblock \bibinfo{title}{{Spectral breaks as a signature of cosmic ray induced turbulence in the Galaxy}}.
\newblock \emph{\bibinfo{journal}{\prl}} \textbf{\bibinfo{volume}{109}}, \bibinfo{pages}{061101} (\bibinfo{year}{2012}).

\bibitem{Shalaby_2023arXiv230518050S}
\bibinfo{author}{{Shalaby}, M.}, \bibinfo{author}{{Thomas}, T.}, \bibinfo{author}{{Pfrommer}, C.}, \bibinfo{author}{{Lemmerz}, R.} \& \bibinfo{author}{{Bresci}, V.}
\newblock \bibinfo{title}{{Deciphering the physical basis of the intermediate-scale instability}}.
\newblock \emph{\bibinfo{journal}{\jpp}} \textbf{\bibinfo{volume}{89}}, \bibinfo{pages}{175890603} (\bibinfo{year}{2023}).

\bibitem{CGL_1956}
\bibinfo{author}{{Chew}, G.~F.}, \bibinfo{author}{{Goldberger}, M.~L.} \& \bibinfo{author}{{Low}, F.~E.}
\newblock \bibinfo{title}{{The Boltzmann equation and the one-fluid hydromagnetic equations in the absence of particle collisions}}.
\newblock \emph{\bibinfo{journal}{Proc. R. Soc. Lond. Ser. A}} \textbf{\bibinfo{volume}{236}}, \bibinfo{pages}{112--118} (\bibinfo{year}{1956}).

\bibitem{Barnes_1966PhFl}
\bibinfo{author}{{Barnes}, A.}
\newblock \bibinfo{title}{{Collisionless damping of hydromagnetic waves}}.
\newblock \emph{\bibinfo{journal}{\pof}} \textbf{\bibinfo{volume}{9}}, \bibinfo{pages}{1483--1495} (\bibinfo{year}{1966}).

\bibitem{Hasegawa_1969PhFl}
\bibinfo{author}{{Hasegawa}, A.}
\newblock \bibinfo{title}{{Drift mirror instability of the magnetosphere.}}
\newblock \emph{\bibinfo{journal}{\pof}} \textbf{\bibinfo{volume}{12}}, \bibinfo{pages}{2642--2650} (\bibinfo{year}{1969}).

\bibitem{Riquelme_2015ApJ...800...27R}
\bibinfo{author}{{Riquelme}, M.~A.}, \bibinfo{author}{{Quataert}, E.} \& \bibinfo{author}{{Verscharen}, D.}
\newblock \bibinfo{title}{{Particle-in-cell simulations of continuously driven mirror and ion cyclotron instabilities in high beta astrophysical and heliospheric plasmas}}.
\newblock \emph{\bibinfo{journal}{\apj}} \textbf{\bibinfo{volume}{800}}, \bibinfo{pages}{27} (\bibinfo{year}{2015}).

\bibitem{Ley_2023arXiv230916751L}
\bibinfo{author}{{Ley}, F.}, \bibinfo{author}{{Zweibel}, E.~G.}, \bibinfo{author}{{Miller}, D.} \& \bibinfo{author}{{Riquelme}, M.}
\newblock \bibinfo{title}{{Secondary whistler and ion-cyclotron instabilities driven by mirror modes in galaxy clusters}}.
\newblock \emph{\bibinfo{journal}{\apj}} \textbf{\bibinfo{volume}{965}}, \bibinfo{pages}{155} (\bibinfo{year}{2024}).

\bibitem{Rosin_2011MNRAS.413....7R}
\bibinfo{author}{{Rosin}, M.~S.}, \bibinfo{author}{{Schekochihin}, A.~A.}, \bibinfo{author}{{Rincon}, F.} \& \bibinfo{author}{{Cowley}, S.~C.}
\newblock \bibinfo{title}{{A non-linear theory of the parallel firehose and gyrothermal instabilities in a weakly collisional plasma}}.
\newblock \emph{\bibinfo{journal}{\mnras}} \textbf{\bibinfo{volume}{413}}, \bibinfo{pages}{7--38} (\bibinfo{year}{2011}).

\bibitem{Hall_1981MNRAS.197..977H}
\bibinfo{author}{{Hall}, A.~N.}
\newblock \bibinfo{title}{{The propagation of Galactic cosmic rays}}.
\newblock \emph{\bibinfo{journal}{\mnras}} \textbf{\bibinfo{volume}{197}}, \bibinfo{pages}{977--993} (\bibinfo{year}{1981}).

\bibitem{Zhou_2023}
\bibinfo{author}{{Zhou}, M.}, \bibinfo{author}{{Zhdankin}, V.}, \bibinfo{author}{{Kunz}, M.~W.}, \bibinfo{author}{{Loureiro}, N.~F.} \& \bibinfo{author}{{Uzdensky}, D.~A.}
\newblock \bibinfo{title}{{Magnetogenesis in a collisionless plasma: from Weibel instability to turbulent dynamo}}.
\newblock \emph{\bibinfo{journal}{\apj}} \textbf{\bibinfo{volume}{960}}, \bibinfo{pages}{12} (\bibinfo{year}{2024}).

\bibitem{Komarov_2016MNRAS.460..467K}
\bibinfo{author}{{Komarov}, S.~V.}, \bibinfo{author}{{Churazov}, E.~M.}, \bibinfo{author}{{Kunz}, M.~W.} \& \bibinfo{author}{{Schekochihin}, A.~A.}
\newblock \bibinfo{title}{{Thermal conduction in a mirror-unstable plasma}}.
\newblock \emph{\bibinfo{journal}{\mnras}} \textbf{\bibinfo{volume}{460}}, \bibinfo{pages}{467--477} (\bibinfo{year}{2016}).

\bibitem{Rincon_2015MNRAS.447L..45R}
\bibinfo{author}{{Rincon}, F.}, \bibinfo{author}{{Schekochihin}, A.~A.} \& \bibinfo{author}{{Cowley}, S.~C.}
\newblock \bibinfo{title}{{Non-linear mirror instability.}}
\newblock \emph{\bibinfo{journal}{\mnras}} \textbf{\bibinfo{volume}{447}}, \bibinfo{pages}{L45--L49} (\bibinfo{year}{2015}).

\bibitem{Melville_2016MNRAS.459.2701M}
\bibinfo{author}{{Melville}, S.}, \bibinfo{author}{{Schekochihin}, A.~A.} \& \bibinfo{author}{{Kunz}, M.~W.}
\newblock \bibinfo{title}{{Pressure-anisotropy-driven microturbulence and magnetic-field evolution in shearing, collisionless plasma}}.
\newblock \emph{\bibinfo{journal}{\mnras}} \textbf{\bibinfo{volume}{459}}, \bibinfo{pages}{2701--2720} (\bibinfo{year}{2016}).

\bibitem{Govoni_2017A&A...603A.122G}
\bibinfo{author}{{Govoni}, F.} \emph{et~al.}
\newblock \bibinfo{title}{{Sardinia Radio Telescope observations of Abell 194. The intra-cluster magnetic field power spectrum}}.
\newblock \emph{\bibinfo{journal}{\aap}} \textbf{\bibinfo{volume}{603}}, \bibinfo{pages}{A122} (\bibinfo{year}{2017}).

\bibitem{Squire_2023arXiv230300468S}
\bibinfo{author}{{Squire}, J.} \emph{et~al.}
\newblock \bibinfo{title}{{Pressure anisotropy and viscous heating in weakly collisional plasma turbulence}}.
\newblock \emph{\bibinfo{journal}{\jpp}} \textbf{\bibinfo{volume}{89}}, \bibinfo{pages}{905890417} (\bibinfo{year}{2023}).

\bibitem{Kalnin_1998JPhA...31.7227K}
\bibinfo{author}{{Kalnin}, J.~R.} \& \bibinfo{author}{{Kotomin}, E.}
\newblock \bibinfo{title}{{Modified Maxwell-Garnett equation for the effective transport coefficients in inhomogeneous media}}.
\newblock \emph{\bibinfo{journal}{J. Phys. A Math. Gen.}} \textbf{\bibinfo{volume}{31}}, \bibinfo{pages}{7227--7234} (\bibinfo{year}{1998}).

\bibitem{Kotera_2008PhRvD..77l3003K}
\bibinfo{author}{{Kotera}, K.} \& \bibinfo{author}{{Lemoine}, M.}
\newblock \bibinfo{title}{{Optical depth of the Universe to ultrahigh energy cosmic ray scattering in the magnetized large scale structure}}.
\newblock \emph{\bibinfo{journal}{\prd}} \textbf{\bibinfo{volume}{77}}, \bibinfo{pages}{123003} (\bibinfo{year}{2008}).

\bibitem{Butsky_2023arXiv230806316B}
\bibinfo{author}{{Butsky}, I.~S.} \emph{et~al.}
\newblock \bibinfo{title}{{Galactic cosmic-ray scattering due to intermittent structures}}.
\newblock \emph{\bibinfo{journal}{\mnras}} \textbf{\bibinfo{volume}{528}}, \bibinfo{pages}{4245--4254} (\bibinfo{year}{2024}).

\bibitem{Arzamasskiy_2023PhRvX..13b1014A}
\bibinfo{author}{{Arzamasskiy}, L.}, \bibinfo{author}{{Kunz}, M.~W.}, \bibinfo{author}{{Squire}, J.}, \bibinfo{author}{{Quataert}, E.} \& \bibinfo{author}{{Schekochihin}, A.~A.}
\newblock \bibinfo{title}{{Kinetic turbulence in collisionless high-{\ensuremath{\beta}} plasmas}}.
\newblock \emph{\bibinfo{journal}{\prx}} \textbf{\bibinfo{volume}{13}}, \bibinfo{pages}{021014} (\bibinfo{year}{2023}).

\bibitem{Majeski_2024arXiv240502418M}
\bibinfo{author}{{Majeski}, S.}, \bibinfo{author}{{Kunz}, M.~W.} \& \bibinfo{author}{{Squire}, J.}
\newblock \bibinfo{title}{{Self-organization in collisionless, high-$\beta$ turbulence}}.
\newblock \emph{\bibinfo{journal}{Journal of Plasma Physics}} \textbf{\bibinfo{volume}{90}}, \bibinfo{pages}{6} (\bibinfo{year}{2024}).

\bibitem{St-Onge2018ApJ...863L..25S}
\bibinfo{author}{{St-Onge}, D.~A.} \& \bibinfo{author}{{Kunz}, M.~W.}
\newblock \bibinfo{title}{{Fluctuation dynamo in a collisionless, weakly magnetized plasma}}.
\newblock \emph{\bibinfo{journal}{\apjl}} \textbf{\bibinfo{volume}{863}}, \bibinfo{pages}{L25} (\bibinfo{year}{2018}).

\bibitem{St-Onge2020JPlPh..86e9003S}
\bibinfo{author}{{St-Onge}, D.~A.}, \bibinfo{author}{{Kunz}, M.~W.}, \bibinfo{author}{{Squire}, J.} \& \bibinfo{author}{{Schekochihin}, A.~A.}
\newblock \bibinfo{title}{{Fluctuation dynamo in a weakly collisional plasma}}.
\newblock \emph{\bibinfo{journal}{\jpp}} \textbf{\bibinfo{volume}{86}}, \bibinfo{pages}{905860503} (\bibinfo{year}{2020}).

\bibitem{Ensslin_2003}
\bibinfo{author}{Enßlin, T.~A.}
\newblock \bibinfo{title}{On the escape of cosmic rays from radio galaxy cocoons}.
\newblock \emph{\bibinfo{journal}{\aap}} \textbf{\bibinfo{volume}{399}}, \bibinfo{pages}{409} (\bibinfo{year}{2003}).

\bibitem{Dursi_Pfrommer_2008}
\bibinfo{author}{Dursi, L.} \& \bibinfo{author}{Pfrommer, C.}
\newblock \bibinfo{title}{Draping of cluster magnetic fields over bullets and bubbles-morphology and dynamic effects}.
\newblock \emph{\bibinfo{journal}{\apj}} \textbf{\bibinfo{volume}{677}}, \bibinfo{pages}{993} (\bibinfo{year}{2008}).

\bibitem{Ruszkowski_2008MNRAS.383.1359R}
\bibinfo{author}{{Ruszkowski}, M.}, \bibinfo{author}{{En{\ss}lin}, T.~A.}, \bibinfo{author}{{Br{\"u}ggen}, M.}, \bibinfo{author}{{Begelman}, M.~C.} \& \bibinfo{author}{{Churazov}, E.}
\newblock \bibinfo{title}{{Cosmic ray confinement in fossil cluster bubbles}}.
\newblock \emph{\bibinfo{journal}{\mnras}} \textbf{\bibinfo{volume}{383}}, \bibinfo{pages}{1359--1365} (\bibinfo{year}{2008}).

\bibitem{Ewart_2024arXiv240405110E}
\bibinfo{author}{{Ewart}, R.~J.}, \bibinfo{author}{{Reichherzer}, P.}, \bibinfo{author}{{Bott}, A. F.~A.}, \bibinfo{author}{{Kunz}, M.~W.} \& \bibinfo{author}{{Schekochihin}, A.~A.}
\newblock \bibinfo{title}{{Cosmic-ray confinement in radio bubbles by micromirrors}}.
\newblock \emph{\bibinfo{journal}{\mnras}} \textbf{\bibinfo{volume}{532}}, \bibinfo{pages}{2098--2107} (\bibinfo{year}{2024}).

\bibitem{Herold2019A&A...625A.110H}
\bibinfo{author}{{Herold}, L.} \& \bibinfo{author}{{Malyshev}, D.}
\newblock \bibinfo{title}{{Hard and bright gamma-ray emission at the base of the Fermi bubbles}}.
\newblock \emph{\bibinfo{journal}{\aap}} \textbf{\bibinfo{volume}{625}}, \bibinfo{pages}{A110} (\bibinfo{year}{2019}).

\bibitem{Giacintucci2024}
\bibinfo{author}{{Giacintucci}, S.}, \bibinfo{author}{{Markevitch}, M.} \& \bibinfo{author}{{Clarke}, T.~E.}
\newblock \bibinfo{title}{{uGMRT observations of the giant fossil radio lobe in the Ophiuchus galaxy cluster}}.
\newblock \emph{\bibinfo{journal}{Rev. Mex. Astron. Astrofis.}} \textbf{\bibinfo{volume}{56}}, \bibinfo{pages}{48--54} (\bibinfo{year}{2024}).

\bibitem{2021NatAs...5..465A}
\bibinfo{author}{{Abeysekara}, A.~U.} \emph{et~al.}
\newblock \bibinfo{title}{{HAWC observations of the acceleration of very-high-energy cosmic rays in the Cygnus Cocoon}}.
\newblock \emph{\bibinfo{journal}{\nastro}} \textbf{\bibinfo{volume}{5}}, \bibinfo{pages}{465--471} (\bibinfo{year}{2021}).

\bibitem{2024SciBu..69..449L}
\bibinfo{author}{{Lhaaso Collaboration}}.
\newblock \bibinfo{title}{{An ultrahigh-energy {\ensuremath{\gamma}} -ray bubble powered by a super PeVatron}}.
\newblock \emph{\bibinfo{journal}{Sci. Bull.}} \textbf{\bibinfo{volume}{69}}, \bibinfo{pages}{449--457} (\bibinfo{year}{2024}).

\bibitem{Miniati_2015}
\bibinfo{author}{{Miniati}, F.}
\newblock \bibinfo{title}{{The Matryoshka Run. II. Time-dependent turbulence statistics, stochastic particle acceleration, and microphysics impact in a massive galaxy cluster}}.
\newblock \emph{\bibinfo{journal}{\apj}} \textbf{\bibinfo{volume}{800}}, \bibinfo{pages}{60} (\bibinfo{year}{2015}).

\bibitem{Brunetti2011MNRAS.412..817B}
\bibinfo{author}{{Brunetti}, G.} \& \bibinfo{author}{{Lazarian}, A.}
\newblock \bibinfo{title}{{Particle reacceleration by compressible turbulence in galaxy clusters: effects of a reduced mean free path}}.
\newblock \emph{\bibinfo{journal}{\mnras}} \textbf{\bibinfo{volume}{412}}, \bibinfo{pages}{817--824} (\bibinfo{year}{2011}).

\bibitem{Brunetti_2016MNRAS.458.2584B}
\bibinfo{author}{{Brunetti}, G.} \& \bibinfo{author}{{Lazarian}, A.}
\newblock \bibinfo{title}{{Stochastic reacceleration of relativistic electrons by turbulent reconnection: a mechanism for cluster-scale radio emission?}}
\newblock \emph{\bibinfo{journal}{\mnras}} \textbf{\bibinfo{volume}{458}}, \bibinfo{pages}{2584--2595} (\bibinfo{year}{2016}).

\bibitem{Bustard_2022ApJ...941...65B}
\bibinfo{author}{{Bustard}, C.} \& \bibinfo{author}{{Oh}, S.~P.}
\newblock \bibinfo{title}{{Turbulent reacceleration of streaming cosmic rays}}.
\newblock \emph{\bibinfo{journal}{\apj}} \textbf{\bibinfo{volume}{941}}, \bibinfo{pages}{65} (\bibinfo{year}{2022}).

\bibitem{Lazarian2023ApJ...956...63L}
\bibinfo{author}{{Lazarian}, A.} \& \bibinfo{author}{{Xu}, S.}
\newblock \bibinfo{title}{{Mirror acceleration of cosmic rays in a high-{\ensuremath{\beta}} medium}}.
\newblock \emph{\bibinfo{journal}{\apj}} \textbf{\bibinfo{volume}{956}}, \bibinfo{pages}{63} (\bibinfo{year}{2023}).

\bibitem{Beduzzi_2023arXiv230603764B}
\bibinfo{author}{{Beduzzi}, L.} \emph{et~al.}
\newblock \bibinfo{title}{{Exploring the origins of mega radio halos}}.
\newblock \emph{\bibinfo{journal}{\aap}} \textbf{\bibinfo{volume}{678}}, \bibinfo{pages}{L8} (\bibinfo{year}{2023}).

\bibitem{Nishiwaki_2024}
\bibinfo{author}{{Nishiwaki}, K.} \& \bibinfo{author}{{Asano}, K.}
\newblock \bibinfo{title}{{Low injection rate of cosmic-ray protons in the turbulent reacceleration model of radio halos in galaxy clusters}}.
\newblock \emph{\bibinfo{journal}{arXiv e-prints}} \bibinfo{pages}{arXiv:2408.13846} (\bibinfo{year}{2024}).

\bibitem{Guo2008MNRAS.384..251G}
\bibinfo{author}{{Guo}, F.} \& \bibinfo{author}{{Oh}, S.~P.}
\newblock \bibinfo{title}{{Feedback heating by cosmic rays in clusters of galaxies}}.
\newblock \emph{\bibinfo{journal}{\mnras}} \textbf{\bibinfo{volume}{384}}, \bibinfo{pages}{251--266} (\bibinfo{year}{2008}).

\bibitem{Jacob_2017MNRAS.467.1449J}
\bibinfo{author}{{Jacob}, S.} \& \bibinfo{author}{{Pfrommer}, C.}
\newblock \bibinfo{title}{{Cosmic ray heating in cool core clusters - I. Diversity of steady state solutions}}.
\newblock \emph{\bibinfo{journal}{\mnras}} \textbf{\bibinfo{volume}{467}}, \bibinfo{pages}{1449--1477} (\bibinfo{year}{2017}).

\bibitem{Jacob_2017MNRAS.467.1478J}
\bibinfo{author}{{Jacob}, S.} \& \bibinfo{author}{{Pfrommer}, C.}
\newblock \bibinfo{title}{{Cosmic ray heating in cool core clusters - II. Self-regulation cycle and non-thermal emission}}.
\newblock \emph{\bibinfo{journal}{\mnras}} \textbf{\bibinfo{volume}{467}}, \bibinfo{pages}{1478--1495} (\bibinfo{year}{2017}).

\bibitem{Cuciti_2022Natur.609..911C}
\bibinfo{author}{{Cuciti}, V.} \emph{et~al.}
\newblock \bibinfo{title}{{Galaxy clusters enveloped by vast volumes of relativistic electrons}}.
\newblock \emph{\bibinfo{journal}{\nat}} \textbf{\bibinfo{volume}{609}}, \bibinfo{pages}{911--914} (\bibinfo{year}{2022}).

\bibitem{Boess_2023MNRAS.519..548B}
\bibinfo{author}{{B{\"o}ss}, L.~M.}, \bibinfo{author}{{Steinwandel}, U.~P.}, \bibinfo{author}{{Dolag}, K.} \& \bibinfo{author}{{Lesch}, H.}
\newblock \bibinfo{title}{{CRESCENDO: an on-the-fly Fokker-Planck solver for spectral cosmic rays in cosmological simulations}}.
\newblock \emph{\bibinfo{journal}{\mnras}} \textbf{\bibinfo{volume}{519}}, \bibinfo{pages}{548--572} (\bibinfo{year}{2023}).

\bibitem{Hopkins_2020MNRAS.492.3465H}
\bibinfo{author}{{Hopkins}, P.~F.} \emph{et~al.}
\newblock \bibinfo{title}{{But what about...: cosmic rays, magnetic fields, conduction, and viscosity in galaxy formation}}.
\newblock \emph{\bibinfo{journal}{\mnras}} \textbf{\bibinfo{volume}{492}}, \bibinfo{pages}{3465--3498} (\bibinfo{year}{2020}).

\bibitem{Ruszkowski_2023arXiv230603141R}
\bibinfo{author}{{Ruszkowski}, M.} \& \bibinfo{author}{{Pfrommer}, C.}
\newblock \bibinfo{title}{{Cosmic ray feedback in galaxies and galaxy clusters}}.
\newblock \emph{\bibinfo{journal}{\aapr}} \textbf{\bibinfo{volume}{31}}, \bibinfo{pages}{4} (\bibinfo{year}{2023}).

\bibitem{Stuardi2021}
\bibinfo{author}{{Stuardi}, C.} \emph{et~al.}
\newblock \bibinfo{title}{{The intracluster magnetic field in the double relic galaxy cluster Abell 2345}}.
\newblock \emph{\bibinfo{journal}{\mnras}} \textbf{\bibinfo{volume}{502}}, \bibinfo{pages}{2518--2535} (\bibinfo{year}{2021}).

\bibitem{Bonafede2010}
\bibinfo{author}{{Bonafede}, A.} \emph{et~al.}
\newblock \bibinfo{title}{{The Coma cluster magnetic field from Faraday rotation measures}}.
\newblock \emph{\bibinfo{journal}{\aap}} \textbf{\bibinfo{volume}{513}}, \bibinfo{pages}{A30} (\bibinfo{year}{2010}).

\bibitem{Steinwandel_2023arXiv230604692S}
\bibinfo{author}{{Steinwandel}, U.~P.}, \bibinfo{author}{{Dolag}, K.}, \bibinfo{author}{{B{\"o}ss}, L.~M.} \& \bibinfo{author}{{Marin-Gilabert}, T.}
\newblock \bibinfo{title}{{Toward cosmological simulations of the magnetized intracluster medium with resolved coulomb collision scale}}.
\newblock \emph{\bibinfo{journal}{\apj}} \textbf{\bibinfo{volume}{967}}, \bibinfo{pages}{125} (\bibinfo{year}{2024}).

\bibitem{2022JCAP...09..035A}
\bibinfo{author}{{Alves Batista}, R.} \emph{et~al.}
\newblock \bibinfo{title}{{CRPropa 3.2 - An advanced framework for high-energy particle propagation in extragalactic and galactic spaces}}.
\newblock \emph{\bibinfo{journal}{\jcap}} \textbf{\bibinfo{volume}{2022}}, \bibinfo{pages}{035} (\bibinfo{year}{2022}).

\bibitem{Mertsch_2020}
\bibinfo{author}{{Mertsch}, P.}
\newblock \bibinfo{title}{{Test particle simulations of cosmic rays}}.
\newblock \emph{\bibinfo{journal}{Astrophysics and Space Science}} \textbf{\bibinfo{volume}{365}}, \bibinfo{pages}{135} (\bibinfo{year}{2020}).

\bibitem{Tautz_2013PhPl...20b2302T}
\bibinfo{author}{{Tautz}, R.~C.} \& \bibinfo{author}{{Dosch}, A.}
\newblock \bibinfo{title}{{On numerical turbulence generation for test-particle simulations}}.
\newblock \emph{\bibinfo{journal}{\pop}} \textbf{\bibinfo{volume}{20}}, \bibinfo{pages}{022302} (\bibinfo{year}{2013}).

\bibitem{Schlegel_2020}
\bibinfo{author}{{Schlegel}, L.}, \bibinfo{author}{{Frie}, A.}, \bibinfo{author}{{Eichmann}, B.}, \bibinfo{author}{{Reichherzer}, P.} \& \bibinfo{author}{{Tjus}, J.~B.}
\newblock \bibinfo{title}{{Interpolation of turbulent magnetic fields and its consequences on cosmic ray propagation}}.
\newblock \emph{\bibinfo{journal}{\apj}} \textbf{\bibinfo{volume}{889}}, \bibinfo{pages}{123} (\bibinfo{year}{2020}).

\bibitem{Giacinti_2012}
\bibinfo{author}{{Giacinti}, G.}, \bibinfo{author}{{Kachelrie{\ss}}, M.}, \bibinfo{author}{{Semikoz}, D.~V.} \& \bibinfo{author}{{Sigl}, G.}
\newblock \bibinfo{title}{{Cosmic ray anisotropy as signature for the transition from galactic to extragalactic cosmic rays}}.
\newblock \emph{\bibinfo{journal}{\jcap}} \textbf{\bibinfo{volume}{2012}}, \bibinfo{pages}{031} (\bibinfo{year}{2012}).

\bibitem{Kempski_2022MNRAS.514..657K}
\bibinfo{author}{{Kempski}, P.} \& \bibinfo{author}{{Quataert}, E.}
\newblock \bibinfo{title}{{Reconciling cosmic-ray transport theory with phenomenological models motivated by Milky-Way data}}.
\newblock \emph{\bibinfo{journal}{\mnras}} \textbf{\bibinfo{volume}{514}}, \bibinfo{pages}{657--674} (\bibinfo{year}{2022}).

\bibitem{Hopkins_2022MNRAS.517.5413H}
\bibinfo{author}{{Hopkins}, P.~F.}, \bibinfo{author}{{Squire}, J.}, \bibinfo{author}{{Butsky}, I.~S.} \& \bibinfo{author}{{Ji}, S.}
\newblock \bibinfo{title}{{Standard self-confinement and extrinsic turbulence models for cosmic ray transport are fundamentally incompatible with observations}}.
\newblock \emph{\bibinfo{journal}{\mnras}} \textbf{\bibinfo{volume}{517}}, \bibinfo{pages}{5413--5448} (\bibinfo{year}{2022}).

\bibitem{Kolmogorov_1941DoSSR..30..301K}
\bibinfo{author}{{Kolmogorov}, A.}
\newblock \bibinfo{title}{{The local structure of turbulence in incompressible viscous fluid for very large Reynolds' numbers}}.
\newblock \emph{\bibinfo{journal}{Dokl. Akad. Nauk SSSR}} \textbf{\bibinfo{volume}{30}}, \bibinfo{pages}{301--305} (\bibinfo{year}{1941}).

\bibitem{Iroshnikov_1964SvA.....7..566I}
\bibinfo{author}{{Iroshnikov}, P.~S.}
\newblock \bibinfo{title}{{Turbulence of a conducting fluid in a strong magnetic field}}.
\newblock \emph{\bibinfo{journal}{SvA}} \textbf{\bibinfo{volume}{7}}, \bibinfo{pages}{566} (\bibinfo{year}{1964}).

\bibitem{Kraichnan_1965PhFl....8.1385K}
\bibinfo{author}{{Kraichnan}, R.~H.}
\newblock \bibinfo{title}{{Inertial-range spectrum of hydromagnetic turbulence}}.
\newblock \emph{\bibinfo{journal}{\pof}} \textbf{\bibinfo{volume}{8}}, \bibinfo{pages}{1385--1387} (\bibinfo{year}{1965}).

\bibitem{GS_1995ApJ...438..763G}
\bibinfo{author}{{Goldreich}, P.} \& \bibinfo{author}{{Sridhar}, S.}
\newblock \bibinfo{title}{{Toward a theory of interstellar turbulence. II. Strong Alfvenic turbulence}}.
\newblock \emph{\bibinfo{journal}{\apj}} \textbf{\bibinfo{volume}{438}}, \bibinfo{pages}{763} (\bibinfo{year}{1995}).

\bibitem{Schekochihin_2022JPlPh..88e1501S}
\bibinfo{author}{{Schekochihin}, A.~A.}
\newblock \bibinfo{title}{{MHD turbulence: a biased review}}.
\newblock \emph{\bibinfo{journal}{\jpp}} \textbf{\bibinfo{volume}{88}}, \bibinfo{pages}{155880501} (\bibinfo{year}{2022}).

\bibitem{Fornieri_2021MNRAS.502.5821F}
\bibinfo{author}{{Fornieri}, O.}, \bibinfo{author}{{Gaggero}, D.}, \bibinfo{author}{{Cerri}, S.~S.}, \bibinfo{author}{{De La Torre Luque}, P.} \& \bibinfo{author}{{Gabici}, S.}
\newblock \bibinfo{title}{{The theory of cosmic ray scattering on pre-existing MHD modes meets data}}.
\newblock \emph{\bibinfo{journal}{\mnras}} \textbf{\bibinfo{volume}{502}}, \bibinfo{pages}{5821--5838} (\bibinfo{year}{2021}).

\bibitem{Chandran_2000PhRvL..85.4656C}
\bibinfo{author}{{Chandran}, B. D.~G.}
\newblock \bibinfo{title}{{Scattering of energetic particles by anisotropic magnetohydrodynamic turbulence with a Goldreich-Sridhar power spectrum}}.
\newblock \emph{\bibinfo{journal}{\prl}} \textbf{\bibinfo{volume}{85}}, \bibinfo{pages}{4656--4659} (\bibinfo{year}{2000}).

\bibitem{Lithwick_2001ApJ...562..279L}
\bibinfo{author}{{Lithwick}, Y.} \& \bibinfo{author}{{Goldreich}, P.}
\newblock \bibinfo{title}{{Compressible magnetohydrodynamic turbulence in interstellar plasmas}}.
\newblock \emph{\bibinfo{journal}{\apj}} \textbf{\bibinfo{volume}{562}}, \bibinfo{pages}{279--296} (\bibinfo{year}{2001}).

\bibitem{Cho_2000ApJ...539..273C}
\bibinfo{author}{{Cho}, J.} \& \bibinfo{author}{{Vishniac}, E.~T.}
\newblock \bibinfo{title}{{The anisotropy of magnetohydrodynamic Alfv{\'e}nic turbulence}}.
\newblock \emph{\bibinfo{journal}{\apj}} \textbf{\bibinfo{volume}{539}}, \bibinfo{pages}{273--282} (\bibinfo{year}{2000}).

\bibitem{Skilling_1975MNRAS.173..255S}
\bibinfo{author}{{Skilling}, J.}
\newblock \bibinfo{title}{{Cosmic ray streaming - III. Self-consistent solutions.}}
\newblock \emph{\bibinfo{journal}{\mnras}} \textbf{\bibinfo{volume}{173}}, \bibinfo{pages}{255--269} (\bibinfo{year}{1975}).

\bibitem{Shukurov_2017ApJ}
\bibinfo{author}{{Shukurov}, A.}, \bibinfo{author}{{Snodin}, A.~P.}, \bibinfo{author}{{Seta}, A.}, \bibinfo{author}{{Bushby}, P.~J.} \& \bibinfo{author}{{Wood}, T.~S.}
\newblock \bibinfo{title}{{Cosmic rays in intermittent magnetic fields}}.
\newblock \emph{\bibinfo{journal}{\apjl}} \textbf{\bibinfo{volume}{839}}, \bibinfo{pages}{L16} (\bibinfo{year}{2017}).

\bibitem{Perri_2019SoPh..294...34P}
\bibinfo{author}{{Perri}, S.}, \bibinfo{author}{{Pucci}, F.}, \bibinfo{author}{{Malara}, F.} \& \bibinfo{author}{{Zimbardo}, G.}
\newblock \bibinfo{title}{{On the power-law distribution of pitch-angle scattering times in solar wind turbulence}}.
\newblock \emph{\bibinfo{journal}{Solar Physics}} \textbf{\bibinfo{volume}{294}}, \bibinfo{pages}{34} (\bibinfo{year}{2019}).

\bibitem{Kadomtsev_1973SPhD...18..115K}
\bibinfo{author}{{Kadomtsev}, B.~B.} \& \bibinfo{author}{{Petviashvili}, V.~I.}
\newblock \bibinfo{title}{{Acoustic turbulence}}.
\newblock \emph{\bibinfo{journal}{Dokl. Phys.}} \textbf{\bibinfo{volume}{18}}, \bibinfo{pages}{115} (\bibinfo{year}{1973}).

\bibitem{Yan_2002PhRvL..89B1102Y}
\bibinfo{author}{{Yan}, H.} \& \bibinfo{author}{{Lazarian}, A.}
\newblock \bibinfo{title}{{Scattering of cosmic rays by magnetohydrodynamic interstellar turbulence}}.
\newblock \emph{\bibinfo{journal}{\prl}} \textbf{\bibinfo{volume}{89}}, \bibinfo{pages}{281102} (\bibinfo{year}{2002}).

\bibitem{Cho_2002PhRvL..88x5001C}
\bibinfo{author}{{Cho}, J.} \& \bibinfo{author}{{Lazarian}, A.}
\newblock \bibinfo{title}{{Compressible sub-Alfv{\'e}nic MHD turbulence in low- {\ensuremath{\beta}} plasmas}}.
\newblock \emph{\bibinfo{journal}{\prl}} \textbf{\bibinfo{volume}{88}}, \bibinfo{pages}{245001} (\bibinfo{year}{2002}).

\bibitem{Tjus_2020PhR...872....1B}
\bibinfo{author}{{Becker Tjus}, J.} \& \bibinfo{author}{{Merten}, L.}
\newblock \bibinfo{title}{{Closing in on the origin of Galactic cosmic rays using multimessenger information}}.
\newblock \emph{\bibinfo{journal}{\physrep}} \textbf{\bibinfo{volume}{872}}, \bibinfo{pages}{1--98} (\bibinfo{year}{2020}).

\bibitem{Hu2020ApJ...901..162H}
\bibinfo{author}{{Hu}, Y.}, \bibinfo{author}{{Lazarian}, A.}, \bibinfo{author}{{Li}, Y.}, \bibinfo{author}{{Zhuravleva}, I.} \& \bibinfo{author}{{Gendron-Marsolais}, M.-L.}
\newblock \bibinfo{title}{{Probing magnetic field morphology in galaxy clusters with the gradient technique}}.
\newblock \emph{\bibinfo{journal}{\apj}} \textbf{\bibinfo{volume}{901}}, \bibinfo{pages}{162} (\bibinfo{year}{2020}).

\bibitem{Brunetti_2007MNRAS.378..245B}
\bibinfo{author}{{Brunetti}, G.} \& \bibinfo{author}{{Lazarian}, A.}
\newblock \bibinfo{title}{{Compressible turbulence in galaxy clusters: physics and stochastic particle re-acceleration}}.
\newblock \emph{\bibinfo{journal}{\mnras}} \textbf{\bibinfo{volume}{378}}, \bibinfo{pages}{245--275} (\bibinfo{year}{2007}).

\bibitem{Krumholz_2020MNRAS.493.2817K}
\bibinfo{author}{{Krumholz}, M.~R.} \emph{et~al.}
\newblock \bibinfo{title}{{Cosmic ray transport in starburst galaxies}}.
\newblock \emph{\bibinfo{journal}{\mnras}} \textbf{\bibinfo{volume}{493}}, \bibinfo{pages}{2817--2833} (\bibinfo{year}{2020}).

\bibitem{Zweibel_2013PhPl...20e5501Z}
\bibinfo{author}{{Zweibel}, E.~G.}
\newblock \bibinfo{title}{{The microphysics and macrophysics of cosmic rays}}.
\newblock \emph{\bibinfo{journal}{\pop}} \textbf{\bibinfo{volume}{20}}, \bibinfo{pages}{055501} (\bibinfo{year}{2013}).

\bibitem{Kempski_2020MNRAS.493.5323K}
\bibinfo{author}{{Kempski}, P.}, \bibinfo{author}{{Quataert}, E.} \& \bibinfo{author}{{Squire}, J.}
\newblock \bibinfo{title}{{Sound-wave instabilities in dilute plasmas with cosmic rays: implications for cosmic ray confinement and the Perseus X-ray ripples}}.
\newblock \emph{\bibinfo{journal}{\mnras}} \textbf{\bibinfo{volume}{493}}, \bibinfo{pages}{5323--5335} (\bibinfo{year}{2020}).

\bibitem{Fabian2006MNRAS.366..417F}
\bibinfo{author}{{Fabian}, A.~C.} \emph{et~al.}
\newblock \bibinfo{title}{{A very deep Chandra observation of the Perseus cluster: shocks, ripples and conduction}}.
\newblock \emph{\bibinfo{journal}{\mnras}} \textbf{\bibinfo{volume}{366}}, \bibinfo{pages}{417--428} (\bibinfo{year}{2006}).

\bibitem{Lyskova2023MNRAS.525..898L}
\bibinfo{author}{{Lyskova}, N.} \emph{et~al.}
\newblock \bibinfo{title}{{X-ray surface brightness and gas density profiles of galaxy clusters up to 3 {\texttimes} R$_{500c}$ with SRG/eROSITA}}.
\newblock \emph{\bibinfo{journal}{\mnras}} \textbf{\bibinfo{volume}{525}}, \bibinfo{pages}{898--907} (\bibinfo{year}{2023}).

\bibitem{chen_renshaw_1992}
\bibinfo{author}{Chen, A.} \& \bibinfo{author}{Renshaw, E.}
\newblock \bibinfo{title}{The gillis–domb–fisher correlated random walk}.
\newblock \emph{\bibinfo{journal}{J. Appl. Probab.}} \textbf{\bibinfo{volume}{29}}, \bibinfo{pages}{792–813} (\bibinfo{year}{1992}).

\bibitem{Bott_2021ApJ...922L..35B}
\bibinfo{author}{{Bott}, A.~F.~A.}, \bibinfo{author}{{Arzamasskiy}, L.}, \bibinfo{author}{{Kunz}, M.~W.}, \bibinfo{author}{{Quataert}, E.} \& \bibinfo{author}{{Squire}, J.}
\newblock \bibinfo{title}{{Adaptive critical balance and firehose instability in an expanding, turbulent, collisionless plasma}}.
\newblock \emph{\bibinfo{journal}{\apjl}} \textbf{\bibinfo{volume}{922}}, \bibinfo{pages}{L35} (\bibinfo{year}{2021}).

\bibitem{Eyink_2013Natur.497..466E}
\bibinfo{author}{{Eyink}, G.} \emph{et~al.}
\newblock \bibinfo{title}{{Flux-freezing breakdown in high-conductivity magnetohydrodynamic turbulence}}.
\newblock \emph{\bibinfo{journal}{\nat}} \textbf{\bibinfo{volume}{497}}, \bibinfo{pages}{466--469} (\bibinfo{year}{2013}).

\bibitem{JHTDB}
\bibinfo{author}{{Aluie}, H.} \emph{et~al.}
\newblock \bibinfo{title}{Johns hopkins turbulence databases, forced mhd turbulence}.
\newblock \bibinfo{howpublished}{\url{https://doi.org/10.7281/T1930RBS}} (\bibinfo{year}{2019}).

\bibitem{Kuhlen_2022arXiv221105881K}
\bibinfo{author}{{Kuhlen}, M.}, \bibinfo{author}{{Phan}, V. H.~M.} \& \bibinfo{author}{{Mertsch}, P.}
\newblock \bibinfo{title}{{Diffusion of relativistic charged particles and field lines in isotropic turbulence}}.
\newblock \emph{\bibinfo{journal}{arXiv e-prints}} \bibinfo{pages}{arXiv:2211.05881} (\bibinfo{year}{2022}).

\bibitem{ZENODO}
\bibinfo{author}{{Reichherzer}, P.}
\newblock \bibinfo{title}{CR diffusion in micromirrors}.
\newblock \emph{\bibinfo{journal}{zenodo}},
\newblock \bibinfo{howpublished}{\url{https://zenodo.org/records/13942018}} (\bibinfo{year}{2024}).

\end{thebibliography}
\end{document}